\documentclass{article}
\usepackage[utf8]{inputenc}

\date{}
\usepackage{bm}
\usepackage{amsmath}
\usepackage{amsfonts}
\usepackage{graphicx}
\usepackage{times}
\usepackage{natbib}
\usepackage{hyperref}
\usepackage{authblk}

\usepackage{multirow}
\usepackage{longtable}
\usepackage{booktabs}
\usepackage{makecell}
\usepackage{enumitem}

\usepackage{color}

\def\Var{\operatorname{Var}}

\def\E{\mbox{E}}

\newtheorem{theorem}{Theorem}
\newtheorem{assumption}{Assumption}
\newtheorem{remark}{Remark}
\newtheorem{lemma}{Lemma}
\newcommand{\n}{\nonumber}

\title{Spatial Regression With Multiplicative Errors, and Its Application With Lidar Measurements}
\author[1]{Hojun You\thanks{hyou3@uh.edu}}
\author[2]{Wei-Ying Wu}
\author[3]{Chae Young Lim}
\author[4]{Kyubaek Yoon}
\author[4]{Jongeun Choi}

\affil[1]{Department of Mathematics, University of Houston}
\affil[2]{Department of Applied Mathematics, National Dong Hwa University}
\affil[3]{Department of Statistics, Seoul National University}
\affil[4]{Department of Mechanical Engineering, Yonsei University}

\begin{document}

\maketitle

\begin{abstract}
   Multiplicative errors in addition to spatially referenced observations often arise in geodetic applications, particularly in surface estimation with light detection and ranging (LiDAR) measurements. However, spatial regression involving multiplicative errors remains relatively unexplored in such applications. In this regard, we present a penalized modified least squares estimator to handle the complexities of a multiplicative error structure while identifying significant variables in spatially dependent observations for surface estimation. The proposed estimator can be also applied to classical additive error spatial regression. By establishing asymptotic properties of the proposed estimator under increasing domain asymptotics with stochastic sampling design, we provide a rigorous foundation for its effectiveness. A comprehensive simulation study confirms the superior performance of our proposed estimator in accurately estimating and selecting parameters, outperforming existing approaches. To demonstrate its real-world applicability, we employ our proposed method, along with other alternative techniques, to estimate a rotational landslide surface using LiDAR measurements. The results highlight the efficacy and potential of our approach in tackling complex spatial regression problems involving multiplicative errors.
\end{abstract}

\section{Introduction}\label{section:introduction}

Multiplicative errors are commonly observed in various fields, including fields such as geodesy, finance, economics, etc. \citep{xu2000least, taylor2003exponential, shi2014adjustment, iyaniwura2019parameter}. In geodesic applications, such as the global positioning system (GPS) and light detection and ranging (LiDAR), the measurement variability is proportional to  the measurements themselves, suggesting the presence of multiplicative errors \citep{seeber2008satellite, shi2014adjustment}. Surface estimation and volume prediction, common in LiDAR applications, typically rely on linear models \citep{shi2014adjustment, mauro2017analysis}. Given the characteristics of geodesic measurements, employing a linear model with multiplicative error proves to be more appropriate.

A linear model with multiplicative errors for LiDAR-type measurements has the following form.
\begin{equation}\label{eq:intro_multi_linear}
    \bm z = \bm X\bm\theta \cdot\bm \varsigma,
\end{equation}
where $\bm z$ is a measured distance vector, $\bm X$ is a covariate matrix, $\bm \theta$ is a set of parameters of interest, and $\varsigma$ is a multiplicative error. To facilitate working with this model, we take a log-transformation on both sides, yielding the more familiar expression:
\begin{equation}\label{eq:intro_add_nonlinear}
    \bm y = \log(\bm X\bm\theta)+\bm\varepsilon,
\end{equation}
where $\bm y = \log(\bm z),\ \bm\varepsilon = \log(\bm\varsigma)$. It is essential to note that the expectation of $\bm\varepsilon$ may not be zero. Existing approaches in LiDAR applications often assume standard additive zero-mean errors instead of accounting for multiplicative errors \citep{xu2000least, shi2014adjustment}. However, such model misspecification may lead to substantial estimation bias \citep{xu2000least} and inconsistency of the least squares (LS) estimator \citep{bhattacharyya1992inconsistent, you2022regularized}. Thus, obtaining a proper estimator under a valid model becomes crucial to avoid undesirable results. 

Another point we consider is spatial correlation of the errors. Prior literature suggests the presence of spatial correlation in LiDAR measurements \citep{mauro2017analysis, mcroberts2018assessing, uss2019estimation} while many related studies work under the assumption of independent errors. To account for both multiplicative errors and spatial correlation, we propose a modified least squares estimator from  \eqref{eq:intro_add_nonlinear}. We also generalize the log function in \eqref{eq:intro_add_nonlinear} to a general nonlinear function to allow more flexibility when deriving theoretical results in Section \ref{section:material_method}.

The literature has extensively explored the asymptotic properties of nonlinear LS estimators. Initially, \cite{jennrich1969asymp} established the consistency and asymptotic normality of the nonlinear LS estimator with independent errors. Subsequently, \cite{wu1981asymptotic} provided necessary conditions for weakly consistent estimators and demonstrated asymptotic properties under weaker assumptions than those in \cite{jennrich1969asymp}. Further advancements by \cite{lai1994stochastic} and \cite{skouras2000strong} considered the LS estimator's asymptotic properties with stochastic nonlinear functions and martingale difference errors. Recent works, like \cite{wang2020least}, have addressed the consistency of the LS estimator under nonlinear models with heteroscedastic errors. However, these previous studies mainly focused on the classical nonlinear regression framework with zero-mean errors, whereas our proposed method deals with the possibility of non-zero mean errors in the model, originated from the multiplicative structure.

Regarding regression models with multiplicative errors, a few studies have been conducted \citep{xu2000least, xu2013adjustment, lim2014operator, shi2014adjustment, you2022regularized}. \cite{xu2000least}, \cite{xu2013adjustment}, and \cite{shi2014adjustment} proposed adjusted LS estimators for \eqref{eq:intro_multi_linear} and investigated estimation biases without providing theoretical results. \cite{lim2014operator} and \cite{you2022regularized} introduced modified least squares estimators and explored their asymptotic properties under independent errors and time correlated errors, respectively. The works of \cite{lim2014operator} and \cite{you2022regularized} served as motivation for our research to work on a LiDAR-type geodetic practice with the modified least squares estimator and develop asymptotic properties under spatially dependent errors.

In LiDAR-type geodetic practices, such as digital terrain modeling, we often need to estimate a surface with coordinate information only \citep{xu2013adjustment, shi2014adjustment, wang2021ridge}. To achieve accurate surface estimation, it is required to consider various functions of the coordinates that may explain the surface, in addition to the coordinates themselves. For instance, in the study of rotational landslides, \cite{shi2014adjustment} used second-order polynomials of the coordinates to model the curved surface shape. However, including more variables than necessary can lead to overfitting and decrease generalization. To address this issue, we adopt a penalized least squares approach that simultaneously estimates the parameters and selects significant variables.

Penalized variable selection methods have garnered significant attention since the introduction of the least absolute shrinkage and selection operator (LASSO) in \cite{tibshirani1996regression}. \cite{fan2001scad} introduced the concept of the oracle property and the smoothly clipped absolute deviation (SCAD) penalty function. Since then, there have been numerous studies on the penalized approach \citep{fan2004nonconcave, zou2006adaptive, yuan2006model, candes2007dantzig, fan2012variable}, but only a few worked under spatial correlation in presence. \cite{wang2009spatial} developed the asymptotic properties of the penalized LS estimator in spatial linear regression. \cite{zhu2010selection} and \cite{chu2011penalized} investigated the asymptotic properties of a penalized maximum likelihood estimator for lattice data and geostatistical data, respectively. \cite{chu2011penalized} also studied an estimator using covariance-tapering to reduce computational costs. \cite{feng2016variable} considered a penalized quasi-likelihood method for binary responses. \cite{liu2018penalized} and \cite{cai2020variable} addressed spatial autoregressive (SAR) models. \cite{liu2018penalized} considered responses following a SAR model and used a quasi-likelihood method while \cite{cai2020variable} focused on LS estimation with SAR disturbances in the model. To the best of our knowledge, our work is the first to take the spatial correlation of the errors into account in a multiplicative regression framework, even without penalization for variable selection. Our proposed approach aims to fill this gap and provide a robust and efficient estimator for surface estimation in LiDAR-type geodetic applications.

The remainder of the article is organized as follows. In Section \ref{section:material_method}, we suggest a modified least squares estimator to account for the possible non-zero mean of the errors. Then, we introduce theorems for the proposed estimator as well as assumptions for the theorems. We investigate the finite sample performance of the proposed estimator and provide results for the LiDAR example in Section \ref{section:results}. At last, we summarize our results and give a brief discussion on our study in Section \ref{section:conclusion}. The detailed proofs are given in the Appendix. 

\section{Materials and methods}\label{section:material_method}

\subsection{Multiplicative regression with spatially correlated errors}

We consider a regression model for a LiDAR system as follows.
\begin{equation}
\label{eq:multiplicative_linear}
    z(\bm s_i)=\bm x(\bm s_i)^T\bm\theta\times\varsigma(\bm s_i)
\end{equation}
for $i \in \{1, \ldots, n\}$. ${A}^{T}$ is the transpose of a matrix $A$. $\bm x(\bm s_i)$ and $\bm \theta_0$ are $p$-dimensional vectors and a random error, $\varsigma(\bm s)$, is spatially correlated and almost surely positive with $\E(\varsigma(\bm s))=1$. The covariates for the LiDAR system are typically locations and functions of the locations \citep{xu2013adjustment, shi2014adjustment, wang2021ridge}. Given $\bm x(\bm s_i)^T\bm\theta_0$ is positive, the model \eqref{eq:multiplicative_linear} is naturally converted to the model \eqref{eq:additive_nonlinear} by the log transformation.
\begin{equation}
    \label{eq:additive_nonlinear}
    y(\bm s_i)= g(\bm x(\bm s_i);\bm\theta)+\varepsilon(\bm s_i),
\end{equation}
where $y(\bm s_i) = \log(z(\bm s_i)),\ g(\bm x(\bm s_i);\bm\theta)=\log(\bm x(\bm s_i)^T\bm\theta)$ and $\varepsilon(\bm s_i) = \log(\varsigma(\bm s_i))$. Differently from a classical nonlinear regression, 
$\E(\varepsilon(\bm s_i))$ may be neither zero nor identified due to confounding with a possible constant term from $g(\cdot; \cdot)$. To remedy this difficulty, we propose a spatial version of the modified least squares \citep{lim2014operator} and \cite{you2022regularized}. Specifically, the modified least squares estimator is obtained by minimizing
\begin{equation}\label{eq:ls}
    S_n(\bm \theta) = \sum_{i=1}^n \left(y(\bm s_i) - g(\bm x(\bm s_i); \bm \theta)-\frac{1}{n}\sum_{j=1}^n  \left(y(\bm s_j) - g(\bm x(\bm s_j); \bm \theta) \right) \right)^2.
\end{equation}
In a squared term, the sample mean of the residuals is subtracted from each residual to exclude possible non-zero mean of the error in the  estimation procedure.

We further assume only a few parameters of the true $\bm \theta_0$ are non-zero and without loss of generality, let the first $s$ entries of the true $\bm\theta_0$ be non-zero. That is,   $\bm\theta_0=(\theta_{01},\theta_{02},\ldots,\theta_{0s},$ $\theta_{0s+1},\ldots,\theta_{0p})^{T}$ and $\theta_{0i}\neq 0$ for $1\leq i\leq s$ and $\theta_{0i}= 0$ for $s+1 \leq i\leq p$. We also select significant parameters by minimizing a penalized objective function $Q_n(\bm \theta)$, which contains an additional penalty function:
\begin{equation}\label{eq:ls_penalty}
    Q_n(\bm \theta) = S_n(\bm \theta) + n \sum_{i=1}^{p} p_{\lambda_n}(|\theta_i|),
\end{equation}
where $p_{\lambda_n}(\cdot)$ is a known penalty function with a tuning parameter $\lambda_n$. The conditions for $p_{\lambda_n}(\cdot)$ and $\lambda_n$ are discussed in Assumption \ref{assumption:penalty}. 

\subsection{Asymptotic framework and sampling designs}

For a spatial asymptotic framework, fixed, mixed, and increasing-domain asymptotic frameworks are considered in the literature \citep{lahiri2003central, stein2012interpolation}. In this article, we work under the increasing-domain asymptotic framework, which assumes the sampling region grows in proportion to the sample size. Extending the work to mixed-domain asymptotics should be straightforward since \cite{lahiri2003central} also provides asymptotic results for mixed-domain. However, we will leave the extension as future work and focus on increasing-domain asymptotic in this article. Define $R_{0}^*$ as an open connected set of $\left(-1/2, 1/2\right]^{d}$ and $R_0$ as a Borel set with $R_{0}^* \subset R_0 \subset \bar{R}_0^*$, where $\bar{A}$ denotes a closure of $A$. The sampling region $R_n$ is inflated by $\eta_n$, scaling factor, from $R_0$ i.e. $R_n = \eta_n R_0$. For the increasing-domain asymptotic framework, $\eta_n^{-d}n \rightarrow 1$. The details of this sampling design are discussed in \cite{lahiri2003central}.

We adopt a stochastic spatial sampling design \citep{lahiri2003central, wang2009spatial} to add generality in sampling locations and derive the asymptotic properties of the estimators. Let $\phi(\bm u)$ be a probability density function on $R_0$ and $\bm U =\{\bm U_n\}_{n\in \mathcal{N}}$ be a set of independent and identically distributed $d$-dimensional random vectors with the probability density function $\phi(\bm u)$. The sampling locations $\{\bm s_n \}_{n\in \mathcal{N}}$ are constructed from the realization of $\{\bm U_n\}_{n\in \mathcal{N}}$ by
$$\bm s_i = \eta_n \bm u_i,$$
where $\bm u_i$'s are the realizations of $\bm U_i$'s. By introducing $\phi(\bm u)$, we can handle data with sampling locations from non-uniform intensity as well as uniformly distributed sampling locations. 

\subsection{Notation and assumptions}

From the following, we do not restrict a nonlinear function in the model \eqref{eq:additive_nonlinear} to the log of a linear function so that our estimation methods is applied to general nonlinear functions. We reformulate $S_n(\bm \theta)$ to the following matrix form:
$$\displaystyle S_n(\bm \theta)=\left(\bm{y}-\bm g(\bm{x}, \bm{\theta})\right)^{T}\bm\Sigma_n\left(\bm{y}-\bm g(\bm{x}, \bm{\theta})\right),$$ 
where $\bm y = (y(\bm s_1), \ldots, y(\bm s_n))^T$, $\bm g(\bm{x}, \bm{\theta})=(g(\bm x(\bm s_1); \bm \theta), \ldots, g(\bm x(\bm s_n); \bm \theta))^T$ and $\bm\Sigma_n = \bm I-n^{-1} \bm 1 \bm 1^T$. $\bm I$ is the identity matrix, and $\bm 1$ is the column vector of 1's. $\bm x(\cdot) \in \mathcal{D} \subset \mathcal{R}^{d_x}$  and $\bm \theta \in \Theta \subset \mathcal{R}^p$. Let $\bm\Sigma_\varepsilon$ and $\lambda_\varepsilon$ denote $\Var(\bm \varepsilon)$ and the maximum eigenvalue of $\bm\Sigma_\varepsilon$, respectively. 

If $g$ is twice differentiable with respect to $\bm\theta$, let $\bm{g}_k=\left(\partial g(\bm x(\bm s_1),\bm\theta)/\partial\theta_k,\ldots,\right.$ $\left.\partial g(\bm x(\bm s_n),\bm \theta)/\partial\theta_k\right)^{T}$ and $\bm{g}_{kl}=\left(\partial^2 \bm g(\bm x(\bm s_1),\bm \theta)/\partial\theta_k\partial\theta_l,\ldots,\partial^2 \bm g(\bm x(\bm s_n),\bm\theta)/\partial\theta_k\partial\theta_l\right)^{T}$. Using $\bm{g}_k$ and $\bm{g}_{kl}$, we define ${\bm{\dot{G}}}(\bm\theta)=(\bm{g}_1,\ldots,\bm{g}_p)$ and $\bm{{\ddot{G}}}(\bm\theta)$=Block($\bm{g}_{kl}$), so $\bm{\dot{G}}(\bm\theta)$ is $n \times p $ matrix whose $k$th column is $\bm{g}_k$ and $\bm{{\ddot{G}}}$ is a $pn\times p$ block matrix whose $(k,l)$th block is $\bm{g}_{kl}$. Convergence of random variables $\{\mathcal{X}_n : n\geq 1\}$ to a random variable $\mathcal{X}$ in probability and to a probability distribution $\mathcal{L}$ in distribution are denoted by $\mathcal{X}_n\overset{p}{\rightarrow}\mathcal{X}$ and $\mathcal{X}_n\overset{d}{\rightarrow}\mathcal{L}$, respectively. We now introduce assumptions to derive theoretical results for the proposed estimators. 

\begin{assumption}\label{assumption:general}
    \begin{enumerate}[label=(\arabic*)]
        \item The nonlinear function $g\in C^2$ on the compact set $\mathcal{D} \times \Theta$, where $C^2$ is the set of twice continuously differentiable functions. The compactness of the domain implies the first and second derivatives are bounded and uniformly continuous.
        \item For any sequence $\{C_n\}$ with $\lim_{n\rightarrow \infty} C_n \searrow 0$, the number of cubes $C_n(\bm i + [0,1)^d),\ \bm i \in \mathcal{Z}^d$ that intersects both $\mathcal{R}_0$ and $\mathcal{R}_0^c$ is $O(C_n^{-(d-1)})$ as $n\rightarrow \infty$. 
        \item The probability density function $\phi(\bm u)$ of $\bm U_1$ is continuous and everywhere positive on $\bar{R}_0$
        \item There exists a $p$-dimensional constant vector $\bm \Delta_g$ and a $p$-dimensional vector-valued function $\bm r_1(\bm h)$ such that as $n \rightarrow \infty$, 
        $$\E_{\bm U_1} \left[ \frac{\partial g(\bm x(\eta_n \bm U_1); \bm\theta_0)}{\partial \bm \theta}\right]\ \rightarrow \ \bm \Delta_g,$$
        $$\displaystyle\int \frac{\partial g(\bm x(\eta_n \bm u + \bm h); \bm\theta_0)}{\partial \bm\theta} \cdot \phi^2(\bm u)d\bm u\ \rightarrow\ \bm r_1(\bm h),$$
        elementwisely.
        \item There exists a $p\times p$ positive definite matrix $\bm \Delta_{g^2}$ and a $p \times p$ matrix-valued function $\bm r_2(\bm h)$ s.t. as $n \rightarrow \infty$ ,
        $$\displaystyle\E_{\bm U_1}\left[\frac{\partial g(\bm x(\eta_n \bm U_1); \bm\theta_0)}{\partial \bm\theta}\left(\frac{\partial g(\bm x(\eta_n \bm U_1); \bm\theta_0)}{\partial \bm\theta}\right)^T \right]\ \rightarrow \ \bm \Delta_{g^2},$$
        $$\displaystyle\int \frac{\partial g(\bm x(\eta_n \bm u + \bm h); \bm\theta_0)}{\partial \bm\theta}\left(\frac{\partial  g(\bm x(\eta_n \bm u); \bm\theta_0)}{\partial \bm\theta}\right)^T \phi^2(\bm u)d\bm u\ \rightarrow\ \bm r_2(\bm h),$$
        elementwisely.
    \end{enumerate}
\end{assumption}

Before we move on to assumptions for spatially correlated errors and a penalty function, we introduce a strong mixing coefficient for a random field $\{\varepsilon(\bm s):\bm s \in \mathcal{R}^d\}$,  \citep{doukhan1994mixing, wang2009spatial},
$$\alpha(a;b) = \sup\{\tilde{\alpha}(D,E):d(D,E)\geq a,\ D, E \in R(b)\}, \hspace{.5cm} a>0,b>0,$$
where $\tilde{\alpha}(D,E) = \sup\{|P(A\cap B)-P(A)P(B)|:A\in \mathcal{F}_{\varepsilon}(D), B\in \mathcal{F}_{\varepsilon}(E) \}$ and $\mathcal{F}_{\varepsilon}(C)$ is the $\sigma$-field induced by $\varepsilon$ on a subset $C \subset \mathcal{R}^d$. A metric $d(D,E) = \inf \{\|\bm s_1-\bm s_2\|_1 : \bm s_1 \in D, \bm s_2 \in E\}$ and $R(b)$ is the collection of all disjoint unions of cubes in $\mathcal{R}^d$ with a total volume not larger than $b$. 

\begin{assumption}\label{assumption:stationary}(Spatially correlated errors)
	\begin{enumerate}[label=(\arabic*)]
	    \item $\lambda_\varepsilon = o(n)$.
	    \item $\varepsilon(\cdot)$ is stationary and $\lambda_\varepsilon=O(1).$
		\item There exists a $\delta>0$ such that 
		$\displaystyle \sup_{i\geq 1}\ \E|\varepsilon(s_i)|^{2+\delta}<\infty$.
	\end{enumerate}
	\vspace*{0.5cm}
    For $\tau_1 \in (0, \infty),\ \tau_2 \in [0, \infty)$, we assume
$$\alpha(a;b) \leq \alpha_1(a)\alpha_2(b)\ \mbox{with}\ \alpha_1(a) = O(a^{-\tau_1})\ \& \ \alpha_2(b) = O(b^{\tau_2}).$$
    \begin{enumerate}[resume, label=(\arabic*)]
        \item $\tau_1 > (\delta+2)d/\delta$.
        \item For $d \geq 2$, $\tau_2 < (\tau_1-d)/(4d)$.
	\end{enumerate}
\end{assumption}

\begin{assumption}
	\label{assumption:penalty}(Penalty function)\\
	The first and second derivatives of a penalty function, $p_{\lambda_{n}}(\cdot)$, are denoted by $q_{\lambda_{n}}(\cdot)$ and $q_{\lambda_{n}}'(\cdot)$. Let $c_n = \max_{1\leq i \leq s} \{|q_{\lambda_n}(|\theta_{0i}|)|\},\ d_n=\max_{1\leq i \leq s}\{|q_{\lambda_{n}}'(|\theta_{0i}|)|\}$.
	\begin{enumerate}	
		\item [(1)] $c_n =O((\lambda_\varepsilon/n)^{1/2})$ and $d_n = o(1)$.
		\item [(2)] $q_{\lambda_{n}}'(\cdot)$ is Lipschitz continuous given $\lambda_n$.
		\item [(3)] $\lambda_n \rightarrow 0,\ n^{1/2}\lambda_n \rightarrow \infty$ as $n\rightarrow \infty$.
		\item [(4)] $\liminf_{n \rightarrow \infty} \liminf_{\theta \rightarrow 0+} \lambda_n^{-1}q_{\lambda_{n}}(\theta)>0$.
		
	\end{enumerate}
\end{assumption}

Assumptions \ref{assumption:general} and \ref{assumption:stationary} contain mild conditions for the domain of data and parameter space, stochastic sampling framework, and the errors. Assumption \ref{assumption:general}-(2) makes the boundary effect negligible \citep{wang2009spatial}. The condition holds for most regions $R_n$ of practical interest such as spheres, ellipsoids, and many nonconvex star-shaped sets in $\mathcal{R}^d$ \citep{lahiri2003central, lahiri2006resampling}. Assumption \ref{assumption:general}-(3) is a highly general condition for $\phi(\bm u)$ and \cite{lahiri2003central} mentions that a weaker version of Assumption \ref{assumption:general}-(3) suffices for the CLT of a weighted sum of the errors. Assumptions \ref{assumption:general}-(4) and (5) together provide a spatial version of the Grenander condition \citep{lahiri2003central, wang2009spatial}. Assumption \ref{assumption:stationary}-(1) guarantees the flexibility of $\bm\Sigma_\varepsilon$ and is necessary for the consistency of the estimators from $S_n(\bm\theta)$ and $Q_n(\bm\theta)$. Assumption \ref{assumption:stationary}-(2), which is stronger than Assumption \ref{assumption:stationary}-(1), in addition to Assumptions \ref{assumption:stationary}-(3)$\sim$(5)  are required to derive the asymptotic normality of the estimators. 
We assume the stationarity of the errors and boundedness of the maximum eigenvalue of the covariance matrix in Assumption \ref{assumption:stationary}-(2). Assumptions \ref{assumption:stationary}-(3), (4), and (5) impose regularity conditions for  moment and strong mixing coefficients. For those readers interested in details of Assumptions \ref{assumption:stationary}-(3), (4) and (5), we refer to \cite{lahiri2003central} and \cite{lahiri2006resampling}. Assumption \ref{assumption:penalty} includes (A.6)-(A.9) in \cite{wang2009spatial}. Assumption \ref{assumption:penalty}-(1) guarantees that there exists a $\sqrt{n/\lambda_\varepsilon}$-consistent estimator of $Q_n(\bm \theta)$ and Assumption \ref{assumption:penalty}-(4) makes the penalized estimator sparse. The details are discussed in \cite{wang2009spatial}. 
\begin{remark}\label{remark:penalty}
    The proper order of $\lambda_n$ for LASSO to satisfy Assumptions \ref{assumption:penalty}-(1) and (2) does not satisfy Assumption \ref{assumption:penalty}-(3) and (4). Therefore, LASSO is not allowed to enjoy the consistency and the asymptotic normality simultaneously. On the contrary, SCAD fulfills all conditions in Assumption \ref{assumption:penalty} with properly chosen $\lambda_n$'s.
\end{remark}

\subsection{Asymptotic properties}

Recall that under the stochastic sampling design, $\bm U$ are random sampling sites, so we describe the asymptotic results given $\bm U$ in this section. First, we construct the consistency of the estimators of $S_n(\bm \theta)$ and $Q_n(\bm \theta)$.
\begin{theorem}\label{theorem:consistency} (Consistency)
    Under Assumptions \ref{assumption:general}-(1),(4),(5) and \ref{assumption:stationary}-(1), for any $\varepsilon>0$ and $a_n = (\lambda_\varepsilon/n)^{1/2}$, there exists a positive constant $C$ such that
    $$P\left(\inf_{\|\bm v\|=C} S_n(\bm \theta_0+a_n\bm v) - S_n(\bm \theta_0) >0 \right) > 1-\varepsilon,$$
    for a large enough $n$. Therefore, with probability tending to 1, there exists a local minimizer of $S_n (\bm \theta)$, $\hat{\bm \theta}^{(s)}$, in the ball centered at $\bm \theta_0$ with the radius $a_n\bm v$. 
\end{theorem}

\begin{theorem}\label{theorem:consistency-penalty} (Consistency with penalty)
    Under assumptions in Theorem \ref{theorem:consistency} and Assumptions \ref{assumption:penalty}-(1) and (2), for any $\epsilon>0$, there exists a positive constant $C$ such that
    $$P\left(\inf_{\|\bm v\|=C} Q_n(\bm\theta_0+b_n \bm v)-Q_n(\bm\theta_0)>0\right)>1-\epsilon,$$
	for a large enough $n$.  
	Consequently, with probability tending to 1, there exists a local minimizer of $Q_n(\bm{\theta})$, $\hat{\bm{\theta}}$, in the ball centered at $\bm{\theta}_0$ with the radius $(a_n+c_n)\bm{v}$.
\end{theorem}

By Assumption \ref{assumption:stationary}-(1), $a_n=o(1)$ and $b_n = a_n+c_n = o(1)$, so we attain the consistency of $\hat{\bm\theta}^{(s)}$ and $\hat{\bm\theta}$. If we additionally assume Assumption \ref{assumption:stationary}-(2), $\hat{\bm\theta}^{(s)}$ and $\hat{\bm\theta}$ become $\sqrt{n}$-consistent estimators of $S_n(\bm\theta)$ and $Q_n(\bm\theta)$, respectively.
Next, we deal with the asymptotic normality of a consistent estimator from $S_n (\bm \theta)$ and the oracle property of a consistent estimator from $Q_n (\bm \theta)$.
\begin{theorem}\label{theorem:normality} (Asymptotic normality)
    Under Assumptions \ref{assumption:general} and \ref{assumption:stationary}, given a consistent estimator of $S_n (\bm \theta)$, $\hat{\bm \theta}^{(s)}$, 
    $$n^{1/2}(\hat{\bf{\bm\theta}}^{(s)}-\bm\theta_{0}) \overset{d}{\rightarrow} N\left(0, \bm \Sigma_g^{-1}\bm \Sigma_\theta\bm \Sigma_g^{-1}\right),$$
    where $\bm \Sigma_\theta=\sigma(0)\bm \Sigma_g  + \int \sigma(\bm h) \bm r(\bm h) d\bm h$, $\bm \Sigma_g = \bm \Delta_{g^2}-\bm \Delta_g \bm \Delta_g^T$, $\sigma(\bm h) = Cov[\varepsilon(\bm s), \varepsilon(\bm s+\bm h)]$, and $\bm r(\bm h) = \bm r_2(\bm h)-\bm \Delta_g(\bm r_1(\bm 0)+\bm r_1(\bm h))^T+\E_{\bm U_1}[\phi(\bm U_1)]\bm \Delta_g\bm \Delta_g^T$.
\end{theorem}
\begin{remark}\label{remark:positivedefinite}
    By Assumption \ref{assumption:general}-(3), the probability density function of $\bm U_1$ is continuous and positive on $\bar{R}_0$, so the probability distribution of $\eta_n \bm U_1$ does not degenerate to a single mass. Since $\bm \Sigma_g$ is the limit of the variance of $\frac{\partial g(\bm x(\eta_n \bm U_1); \bm \theta_0)}{\partial \bm \theta}$ by definition, it is a positive definite matrix.
\end{remark}

\begin{theorem}\label{theorem:sparsity} (Oracle property)
     Under the assumptions in Theorem \ref{theorem:normality} and Assumption \ref{assumption:penalty}, given a consistent estimator of $Q_n (\bm \theta)$, $\hat{\bm \theta}$,
     \begin{itemize}
		\item[(i)] for $i \in \{s+1, \ldots, p\}$,
		$P(\hat{\theta}_i= 0)~\rightarrow~ 1$
		\item[(ii)] $n^{1/2}({\hat{\bm \theta}}_{1}-\bm\theta_{01}+(2\bm \Sigma_g)_{11}^{-1}\bm \beta_{n,s}) ~\stackrel{d}{\rightarrow}~ N\left(0, \left(\bm\Sigma_g^{-1}\bm\Sigma_\theta\bm\Sigma_g^{-1}\right)_{11}\right)$,
	\end{itemize}
	where $\bm \theta_1 = (\theta_1, \ldots, \theta_s)^T,\ \bm \beta_{n,s}=(q_{\lambda_{n}}({|\theta_{01}|)sgn(\theta}_{01}),\ldots,q_{\lambda_{n}}(|\theta_{0s}|)sgn(\theta_{0s}))^{T}$ and $\bm A_{11}$ is the $s\times s$ upper-left matrix of $\bm A$.
\end{theorem}
\section{Numerical results}\label{section:results}

In this section, we investigate finite sample performance of the penalized estimator from $Q_n(\bm\theta)$ and show estimation results from a LiDAR application.

\subsection{Simulation study}

Now, we explore finite sample performance of the penalized estimator we proposed with two well-known penalty functions, LASSO and SCAD \citep{tibshirani1996regression, fan2001scad}. First, we generate data from a nonlinear additive regression:
\begin{equation}\label{eq:simulation}
    y_i = \frac{1}{1+\theta_{01} \exp{(-\bm\theta_{0, 2:p}^T\bm x_i)} }+\epsilon_i,
\end{equation}
where $\bm \theta_0 = (\theta_{01}, \theta_{02}, \ldots, \theta_{0p})^T=(1,4,3,2,1,0,\ldots,0)^T$ and $\bm\theta_{0, i:p}=(\theta_{0i}, \ldots, \theta_{0p})^T$ with $p=20$. Be noted that the mean of $\epsilon_i$ may not be zero from \eqref{eq:multiplicative_linear} and \eqref{eq:additive_nonlinear}. The true parameter configuration refers to \cite{wang2009spatial} and the sample size increases from $n=50$ to $200$. The covariates are sampled from a multivariate Gaussian distribution with zero mean, unit variance and correlation of 0.5. For the error, $\epsilon_i$, we simulate a Gaussian random field on $[0,1]\times[0,1]$ with non-zero mean $\mu\in \{0.1, 0.5\}$ and standard deviation $\sigma \in \{0.1, 0.5\}$. 
 The sampling domain increases proportionally to the square root of the sample size since we are working under the increasing domain framework. Two auto-covariance functions (exponential and Gaussian) with two range parameter values ($\rho\in \{1, 2\}$) and the nugget effect of $0.2$ are considered.

We implement a cyclic coordinate descent (CD) algorithm for optimization. The CD algorithms we used for LASSO and non-convex penalty functions including SCAD refer to \cite{wu2008coordinate} and \cite{breheny2011coordinate}, respectively. Our proposed method is denoted as penalized modified least squares (PMLS) and an alternative is denoted as penalized ordinary least squares (POLS). POLS introduces an intercept for estimating the non-zero mean of the error and regards the error has the zero mean. So, POLS minimizes
\begin{equation}\label{eq:POLS}
(\bm y - \beta_0-\bm g(\bm x; \bm\theta))^T(\bm y - \beta_0-\bm g(\bm x; \bm\theta))+n \sum_{j=1}^{p}p_{\lambda_n}(|\theta_j|),
\end{equation}
where $\beta_0$ is the intercept term.

The tuning parameter $\lambda_n$ is chosen by minimizing a BIC-type criterion \citep{wang2007tuning}. The BIC-type criterion is defined as $\text{BIC} = \log(\hat{\sigma}^2)+\log(n)\cdot \widehat{df}/n$, where $\hat{\sigma}^2$ is an estimate for $\sigma^2$ and $\widehat{df}$ is the number of significant estimates. We use $\hat{\sigma}^2=\overline{r^2}-\overline{r}^2$ where $\overline{r}=n^{-1}\sum_{i=1}^{n}r_i$ and $\overline{r^2}=n^{-1}\sum_{i=1}^{n}r_i^2$ and $r_i = y(\bm s_i)-g(\bm x(\bm s_i); \hat{\bm\theta})$. An additional tuning parameter in SCAD is fixed at 3.7 as recommended in \cite{fan2001scad}.

\begin{table}
	\small
	\centering
	\begin{tabular}{cllrrr}
	\Xhline{3\arrayrulewidth}
    $(\mu,\sigma)$ & Cov model & Method & \multicolumn{1}{c}{$n=50$} & \multicolumn{1}{c}{$n=100$} & \multicolumn{1}{c}{$n=200$} \\ \hline
	
	\multirow{8}{*}{$(0.1,0.5)$} & \multirow{2}{*}{$\mbox{Exp}_1$} & PMLS & 23.55 (1.51) & 14.67 (1.30) & 2.76 (0.57) \\ \cline{3-3}
	& & POLS & 19.69 (1.49) & 10.69 (1.16) & 4.31 (0.82) \\\cline{2-3}
	& \multirow{2}{*}{$\mbox{Exp}_2$} & PMLS & 14.75 (1.24) & 4.55 (0.74) & 2.60 (0.56) \\ \cline{3-3}
	& & POLS & 14.89 (1.34) & 6.93 (0.94) & 2.68 (0.67) \\\cline{2-3}
	& \multirow{2}{*}{$\mbox{Gauss}_1$} & PMLS & 16.32 (1.29) & 9.66 (1.09) & 2.38 (0.53) \\ \cline{3-3}
	& & POLS & 13.16 (1.29) & 8.38 (1.08) & 2.61 (0.67) \\\cline{2-3}
	& \multirow{2}{*}{$\mbox{Gauss}_2$} & PMLS & 11.08 (1.16) & 3.65 (0.65) & 1.13 (0.35) \\\cline{3-3}
	& & POLS & 8.71 (1.09) & 4.18 (0.82) & 1.03 (0.52) \\\hline
	
	\multirow{8}{*}{$(0.5,0.5)$} & \multirow{2}{*}{$\mbox{Exp}_1$} & PMLS & 19.46 (1.35) & 11.71 (1.20) & 4.55 (0.75) \\ \cline{3-3}
	& & POLS & 16.71 (1.36) & 8.73 (1.05) & 5.78 (0.96) \\\cline{2-3}
	& \multirow{2}{*}{$\mbox{Exp}_2$} & PMLS & 20.09 (1.43) & 9.54 (1.09) & 2.99 (0.63) \\ \cline{3-3}
	& & POLS & 14.02 (1.36) & 5.98 (0.98) & 3.57 (0.80) \\\cline{2-3}
	& \multirow{2}{*}{$\mbox{Gauss}_1$} & PMLS & 18.24 (1.42) & 6.89 (0.93) & 2.50 (0.53) \\ \cline{3-3}
	& & POLS & 13.30 (1.26) & 7.24 (1.02) & 3.80 (0.78) \\\cline{2-3}
	& \multirow{2}{*}{$\mbox{Gauss}_2$} & PMLS & 18.23 (1.39) & 4.54 (0.75) & 1.93 (0.48) \\\cline{3-3}
	& & POLS & 8.29 (1.07) & 2.31 (0.65) & 1.54 (0.56) \\
	\Xhline{3\arrayrulewidth}
	\multicolumn{5}{l}{\footnotesize{$\ast$ The actual MSE values are $0.01 \times$ the reported values.}}
	\end{tabular}
    \caption{Estimation results with LASSO from \eqref{eq:simulation}. The error is generated from a multivariate Gaussian distribution with $\mu$ (mean) and $\sigma$ (standard deviation). A covariance model with a subscript indicates a corresponding auto-covariance function with a value of the range parameter ($1$ or $2$). The reported values are MSE with SD in the parenthesis.}
    \label{table:lasso_estimation}
\end{table}
\begin{table}
	\small
	\centering
	\begin{tabular}{cllrrr}
	\Xhline{3\arrayrulewidth}
    $(\mu,\sigma)$ & Cov model & Method & \multicolumn{1}{c}{$n=50$} & \multicolumn{1}{c}{$n=100$} & \multicolumn{1}{c}{$n=200$} \\ \hline
	
	\multirow{8}{*}{$(0.1,0.5)$} & \multirow{2}{*}{$\mbox{Exp}_1$} & PMLS & 7.45 (1.21) & 6.24 (1.12) & 2.86 (0.75) \\ \cline{3-3}
	& & POLS & 19.80 (1.98) & 7.85 (1.26) & 4.25 (0.93) \\\cline{2-3}
	& \multirow{2}{*}{$\mbox{Exp}_2$} & PMLS & 5.97 (1.08) & 4.80 (0.98) & 2.50 (0.70) \\ \cline{3-3}
	& & POLS & 7.14 (1.19) & 4.04 (0.93) & 2.88 (0.76) \\\cline{2-3}
	& \multirow{2}{*}{$\mbox{Gauss}_1$} & PMLS & 5.90 (1.06) & 4.09 (0.91) & 2.75 (0.74)  \\ \cline{3-3}
	& & POLS & 6.21 (1.11) & 4.14 (0.94) & 3.64 (0.87) \\\cline{2-3}
	& \multirow{2}{*}{$\mbox{Gauss}_2$} & PMLS & 4.17 (0.89) & 3.75 (0.87) & 1.84 (0.61) \\\cline{3-3}
	& & POLS & 6.79 (1.18) & 3.72 (0.91) & 3.18 (0.86) \\\hline
	
	\multirow{8}{*}{$(0.5,0.5)$} & \multirow{2}{*}{$\mbox{Exp}_1$} & PMLS & 10.30 (1.42) & 4.80 (0.98) & 3.10 (0.79) \\ \cline{3-3}
	& & POLS & 8.68 (1.36) & 4.83 (1.04) & 3.35 (0.89) \\\cline{2-3}
	& \multirow{2}{*}{$\mbox{Exp}_2$} & PMLS & 7.31 (1.20) & 3.68 (0.85) & 2.18 (0.66) \\ \cline{3-3}
	& & POLS & 6.30 (1.18) & 3.74 (0.94) & 2.91 (0.84) \\\cline{2-3}
	& \multirow{2}{*}{$\mbox{Gauss}_1$} & PMLS & 8.49 (1.29) & 4.35 (0.93) & 2.13 (0.65) \\ \cline{3-3}
	& & POLS & 6.26 (1.17) & 2.73 (0.83) & 1.43 (0.65) \\\cline{2-3}
	& \multirow{2}{*}{$\mbox{Gauss}_2$} & PMLS & 5.83 (1.07) & 2.98 (0.76) & 1.13 (0.47) \\\cline{3-3}
	& & POLS & 10.06 (1.48) & 7.40 (1.29) & 6.98 (1.24) \\
	\Xhline{3\arrayrulewidth}
	\multicolumn{5}{l}{\footnotesize{$\ast$ The actual MSE values are $0.01 \times$ the reported values.}}
	\end{tabular}
    \caption{Estimation results with SCAD from \eqref{eq:simulation}. The other configurations are identical to Table \ref{table:lasso_estimation}.}
    \label{table:scad_estimation}
\end{table}
\begin{table}
	\small
	\centering
	\begin{tabular}{cllrrrrrrr}
	\Xhline{3\arrayrulewidth}
    \multirow{2}{*}{$(\mu,\sigma)$} & \multirow{2}{*}{Cov model} & \multirow{2}{*}{Method} & \multicolumn{3}{c}{TP} & & \multicolumn{3}{c}{TN} \\ \cline{4-6} \cline{8-10}
     & & & $50$ & $100$ & $200$ & & $50$ & $100$ & $200$ \\ \hline
	
	\multirow{8}{*}{$(0.1,0.5)$} & \multirow{2}{*}{$\mbox{Exp}_1$} & PMLS & 4.62 & 4.83 & 4.99 & & 14.47 & 13.74 & 12.93 \\ \cline{3-3}
	& & POLS & 4.62 & 4.86 & 4.97 & & 14.52 & 14.00 & 13.77 \\\cline{2-3}
	& \multirow{2}{*}{$\mbox{Exp}_2$} & PMLS & 4.82 & 4.95 & 4.99 & & 14.00 & 13.26 & 13.14 \\ \cline{3-3}
	& & POLS & 4.75 & 4.91 & 4.99 & & 14.50 & 13.96 & 14.02 \\\cline{2-3}
	& \multirow{2}{*}{$\mbox{Gauss}_1$} & PMLS & 4.76 & 4.90 & 4.99 & & 14.24 & 13.44 & 12.98 \\ \cline{3-3}
	& & POLS & 4.78 & 4.90 & 4.99 & & 14.49 & 13.85 & 13.66 \\\cline{2-3}
	& \multirow{2}{*}{$\mbox{Gauss}_2$} & PMLS & 4.88 & 4.96 & 5.00 & & 13.84 & 13.21 & 12.35 \\\cline{3-3}
	& & POLS & 4.87 & 4.95 & 5.00 & & 14.17 & 14.08 & 12.98 \\\cline{2-3}
	\hline
	
	\multirow{8}{*}{$(0.5,0.5)$} & \multirow{2}{*}{$\mbox{Exp}_1$} & PMLS & 4.74 & 4.85 & 4.98 & & 14.37 & 13.82 & 12.97 \\ \cline{3-3}
	& & POLS & 4.69 & 4.90 & 4.96 & & 14.65 & 13.84 & 13.00 \\\cline{2-3}
	& \multirow{2}{*}{$\mbox{Exp}_2$} & PMLS & 4.74 & 4.93 & 4.98 & & 14.29 & 13.76 & 12.81 \\ \cline{3-3}
	& & POLS & 4.84 & 4.95 & 4.96 & & 14.32 & 13.56 & 13.14 \\\cline{2-3}
	& \multirow{2}{*}{$\mbox{Gauss}_1$} & PMLS & 4.76 & 4.94 & 4.98 & & 14.01 & 13.24 & 13.32 \\ \cline{3-3}
	& & POLS & 4.79 & 4.92 & 4.99 & & 14.47 & 13.70 & 13.25 \\\cline{2-3}
	& \multirow{2}{*}{$\mbox{Gauss}_2$} & PMLS & 4.78 & 4.98 & 5.00 & & 14.00 & 12.68 & 12.69 \\\cline{3-3}
	& & POLS & 4.90 & 4.98 & 5.00 & & 14.30 & 13.11 & 12.79 \\
	\Xhline{3\arrayrulewidth}
	\end{tabular}
    \caption{Selection results with LASSO from \eqref{eq:simulation}. TP counts the number of significant estimates among the significant true parameters and TN counts the number of insignificant estimates among the insignificant true parameters. The other configurations are identical to Table \ref{table:lasso_estimation}.}
    \label{table:lasso_selection}
\end{table}

\begin{table}
	\small
	\centering
	\begin{tabular}{cllrrrrrrr}
	\Xhline{3\arrayrulewidth}
    \multirow{2}{*}{$(\mu,\sigma)$} & \multirow{2}{*}{Cov model} & \multirow{2}{*}{Method} & \multicolumn{3}{c}{TP} & & \multicolumn{3}{c}{TN} \\ \cline{4-6} \cline{8-10}
     & & & $50$ & $100$ & $200$ & & $50$ & $100$ & $200$ \\ \hline
	
	\multirow{8}{*}{$(0.1,0.5)$} & \multirow{2}{*}{$\mbox{Exp}_1$} & PMLS & 4.74 & 4.74 & 4.81 & & 15.00 & 15.00 & 15.00 \\ \cline{3-3}
	& & POLS & 4.80 & 4.86 & 4.91 & & 14.23 & 14.59 & 14.42 \\\cline{2-3}
	& \multirow{2}{*}{$\mbox{Exp}_2$} & PMLS & 4.76 & 4.77 & 4.85 & & 14.99 & 14.98 & 15.00 \\ \cline{3-3}
	& & POLS & 4.82 & 4.91 & 4.95 & & 14.37 & 14.79 & 14.76 \\\cline{2-3}
	& \multirow{2}{*}{$\mbox{Gauss}_1$} & PMLS & 4.71 & 4.81 & 4.87 & & 15.00 & 15.00 & 15.00 \\ \cline{3-3}
	& & POLS & 4.80 & 4.87 & 4.94 & & 14.52 & 14.72 & 14.60 \\\cline{2-3}
	& \multirow{2}{*}{$\mbox{Gauss}_2$} & PMLS & 4.79 & 4.80 & 4.92 & & 15.00 & 15.00 & 15.00 \\\cline{3-3}
	& & POLS & 4.81 & 4.93 & 4.97 & & 14.23 & 14.52 & 14.34 \\\cline{2-3}
	\hline
	
	\multirow{8}{*}{$(0.5,0.5)$} & \multirow{2}{*}{$\mbox{Exp}_1$} & PMLS & 4.65 & 4.72 & 4.80 & & 14.99 & 15.00 & 15.00 \\ \cline{3-3}
	& & POLS & 4.87 & 4.94 & 4.96 & & 14.84 & 14.85 & 14.85 \\\cline{2-3}
	& \multirow{2}{*}{$\mbox{Exp}_2$} & PMLS & 4.70 & 4.77 & 4.88 & & 14.98 & 15.00 & 15.00 \\ \cline{3-3}
	& & POLS & 4.91 & 4.95 & 4.97 & & 14.85 & 14.85 & 14.85\\\cline{2-3}
	& \multirow{2}{*}{$\mbox{Gauss}_1$} & PMLS & 4.68 & 4.75 & 4.88 & & 14.98 & 15.00 & 15.00 \\ \cline{3-3}
	& & POLS & 4.89 & 4.94 & 4.97 & & 15.00 & 14.97 & 15.00 \\\cline{2-3}
	& \multirow{2}{*}{$\mbox{Gauss}_2$} & PMLS & 4.70 & 4.79 & 4.98 & & 14.98 & 15.00 & 14.97 \\\cline{3-3}
	& & POLS & 4.85 & 4.95 & 4.98 & & 14.40 & 14.40 & 14.40 \\
	\Xhline{3\arrayrulewidth}
	\end{tabular}
    \caption{Selection results with SCAD from \eqref{eq:simulation}. The other configurations are identical to Table \ref{table:lasso_estimation}.}
    \label{table:scad_selection}
\end{table}

The results of 100 repetitions with LASSO and SCAD are described in Tables \ref{table:lasso_estimation}- \ref{table:scad_selection}. We only report when $\sigma=0.5$ since it is a more challenging case. 
 Mean squared error (MSE), standard deviation (SD), true positive (TP) and true negative (TN) are the criteria we use to examine the finite sample performance of the estimators. MSE and SD are calculated as
\begin{align*}
    MSE & = \frac{1}{Rp}\sum_{j=1}^{R}\sum_{i=1}^{p}\left(\hat\theta_i^{(j)}-\theta_{0,i}\right)^2, \\
    SD & = \displaystyle \sqrt{\frac{1}{R-1} \sum_{j=1}^{R} \left\|\hat{\bm\theta}^{(j)}-{\bm{\bar\theta}}\right\|_2^2},
\end{align*}
where $\hat\theta_i^{(j)}$ stands for the $i$-th component of the estimate from the $j$-th repetition and $p$ and $R$ represent the number of parameters and repetitions, respectively. TP counts the number of significant estimate components among the significant true parameter components and TN counts the number of insignificant estimates among the insignificant true parameter components.

Tables \ref{table:lasso_estimation} and \ref{table:scad_estimation} contain estimation results with LASSO and SCAD, respectively. In Table \ref{table:lasso_estimation}, PMLS with LASSO yields accurate estimates even with relatively small sample sizes and MSE and SD values become smaller as the sample size grows. POLS also exhibits good estimation performance as well and comparable results with PMLS for $n=100$. However, PMLS outperforms POLS for most auto-covariance models and the range parameters when $n=200$. Similar results can be found with SCAD penalty function in Table \ref{table:scad_estimation}. PMLS and POLS show comparable results with small sample sizes, but as the sample size grows, PMLS is superior to POLS in estimating the parameters. The improved performance of the PMLS estimator compared to the POLS estimator indeed arises from the fact that PMLS estimates one less parameter, specifically the intercept, than POLS. By estimating one less parameter with the same amount of data, PMLS achieves more efficient estimation results.

Tables \ref{table:lasso_selection} and \ref{table:scad_selection} contain selection results with LASSO and SCAD, respectively. In both Tables, TP values for PMLS are very close to the true value, 5, and they approach closer to the true value as the sample size grows. Hence, the results with LASSO exhibit a descent finite sample performance in terms of MSE and TP. However, TN values with LASSO does not approach to the true value, 15, even with the relatively large sample size. Since LASSO does not satisfy Assumption \ref{assumption:penalty} as stated in Remark \ref{remark:penalty}, the result of Theorem \ref{theorem:sparsity} supports the behavior of the TN values. On the other hand, we observe TN values for SCAD are almost identical to the true value regardless of the sample sizes. Since Theorem \ref{theorem:sparsity} holds for SCAD with properly chosen $\lambda_n$, TN values for SCAD must converge to the true value and the finite sample performances align well with our expectation. Compared to POLS, PMLS with SCAD displays slightly better accuracy than POLS in terms of TN values. PMLS with LASSO, on the other hand, shows slightly less accuracy than POLS in terms of TN, but the differences may not be statistically significant.  

\subsection{Data example}

\begin{figure}[!ht]
    \centering
    \includegraphics[scale=0.4]{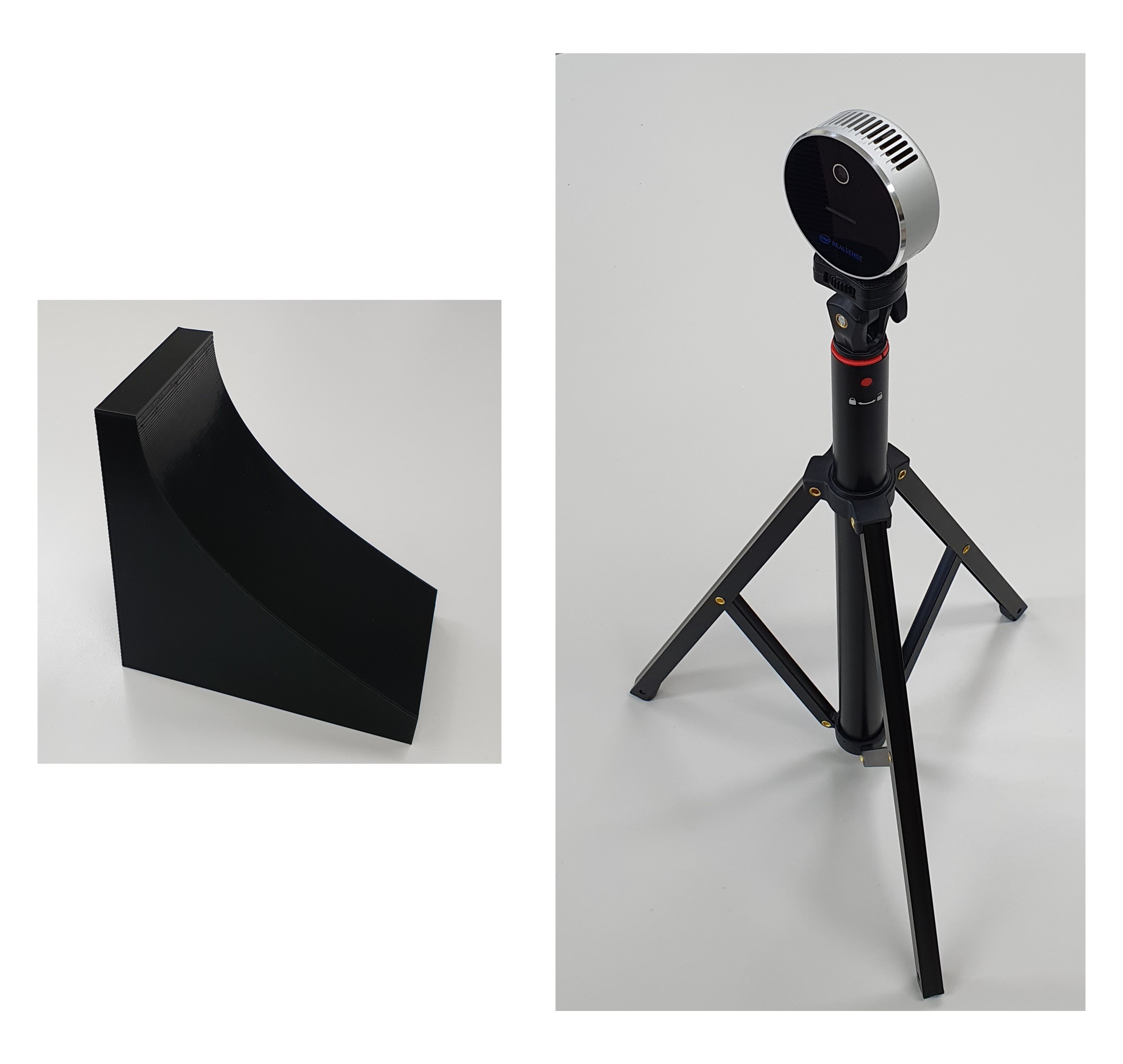}
    \caption{Pictures of the generated object (left) and the LiDAR camera (right). The object is generated by a 3d printer to have curved surface at its upper side where the curve is formed by \eqref{eq:true_curve}. The size of the object is $16cm \times 10cm \times 16cm$. The LiDAR camera is Intel RealSense LiDAR Camera L515, which measures $640\times480$ points simultaneously. The LiDAR camera is set up on a tripod so that the camera is located at the same height as the curved object on a table.}
    \label{fig:lidar_objects}
\end{figure}

\begin{figure}[!ht]
    \centering
    \includegraphics[scale=0.5]{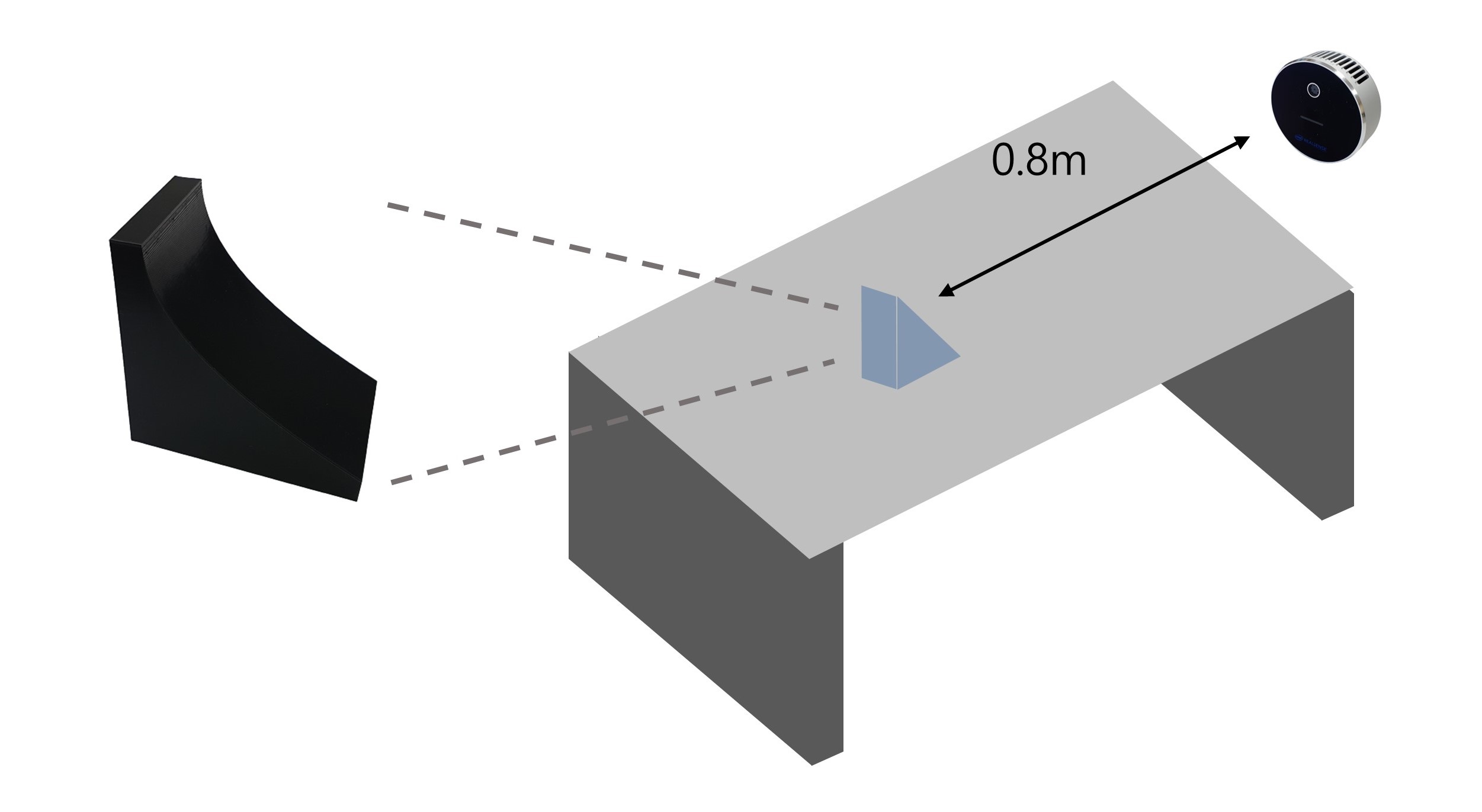}
    \caption{The illustration of the experiment. The curved object is put on a table at the same height as a LiDAR camera. The camera faces the curved surface of the object to the front at the distance of $0.8$ meter.}
    \label{fig:experiment}
\end{figure}

In this section, we explain LiDAR applications and obtain the proposed modified LS estimates with the SCAD penalty. LiDAR measurements have been verified to display multiplicative error structure both in theoretical and practical aspects \citep{xu2013adjustment, shi2014adjustment, shi2020adjustment}. Thus, we create a LiDAR measurement example to examine the performance of our proposed method. We slightly modified a data example given in \cite{shi2014adjustment} about a rotational landslide model. A rotational landslide often leaves a curved surface and LiDAR-type measurements are utilized for estimating the surface. Thus, we aim to estimate the surface function with $x$ and $y$ coordinates using LiDAR measurements in this example. We generated an object with a rotational landslide surface by a 3d printer (Sindoh-3DWOX). Following \cite{shi2014adjustment}, we created the curved surface from a quadratic function as follows. 
\begin{equation}\label{eq:true_curve}
    z = 0.07(y-2)^2-1.98(y-2)+16,
\end{equation}
The quadratic function is designed to pass $(2,16)$ and $(16,2)$ and the surface of the object is the curved line between $(2,16)$ and $(16,2)$. The object has a size of $16\text{cm} \times 10\text{cm} \times 16\text{cm}$. The object and the experiment are described in Figure \ref{fig:lidar_objects} and \ref{fig:experiment}, respectively. 

We used Intel RealSense LiDAR Camera L515 (\url{https://www.intelrealsense.com/lidar-camera-l515/}) to measure the surface. We place the object on a table and set the LiDAR camera on a tripod with the same height as the table where the object is located 0.8 meter away from the camera. The object is positioned so that the curved surface faces the LiDAR camera, then the LiDAR camera measures depths (negative distance) from the surface. The output resolution of the LiDAR camera is 640$\times$480, which means it measures distances from 640$\times$480 points at once. At last, we slice the area where the curved surface is located out of the 640$\times$480 points, which leaves 682 observations in total. This experiment is a simulated example of the LiDAR-type measurements, so this cannot represent all complexity of real applications. The insights gained from this experiment can be transferred to practical geodetic applications that utilize LiDAR measurements for surface estimation or volume prediction, offering valuable guidance to practitioners in related fields.

Although the surface is generated by \eqref{eq:true_curve}, when the LiDAR camera observes the surface, the equation is converted to \eqref{eq:lidar_curve} due to the positional relationship between the camera and the object.
\begin{equation}\label{eq:lidar_curve}
    z = 0.07\cdot y^2+a\cdot y+b,
\end{equation}
where $z$ is a LiDAR measurement, $y$ is a $y$-coordinate value, $a$ and $b$ cannot be specified since their values differ from the locations of the object and the LiDAR camera. Note that the curve function does not depend on a $x$ coordinate. Therefore, our results will focus on selection performance and the coefficient of $y^2$.

The distance response ($z$), $x$ and $y$ coordinates are recorded from the LiDAR camera. Our regression model for $z$ is a second-order polynomial as follows.
$$z = \left(\theta_1+\theta_2 x +\theta_3 y+\theta_4 x^2+ \theta_5 y^2+\theta_6 xy\right)\cdot \varsigma, $$
where $\varsigma$ is a spatially correlated multiplicative error. Our proposed method and two alternative are considered for comparison. A classical approach considers $\varsigma$ as an additive error (an additive method) while our method (PMLS) takes log-transformation at both sides and minimizes \eqref{eq:ls_penalty} with the SCAD penalty. The second alternative is POLS from \eqref{eq:POLS} for comparison. The additive method is implemented by \textit{ncvfit} function from \textit{ncvreg} package in \textit{R}. The \textit{ncvfit} function fits nonconvex penalized regression and supports optimization with the SCAD penalty. The tuning parameter is chosen by a BIC-type criterion as the simulation study. The results are shown in Table \ref{table:lidar_results}.

\begin{table}[!ht]
    \centering
    \begin{tabular}{c|c c c c c c}
    \Xhline{2\arrayrulewidth}
        Method & Intercept & $x$ & $y$ & $x^2$ & $y^2$ & $xy$ \\ \hline
        Additive & -14.03 & -0.04 & -0.90 & 0 & 0.07 & 0.0035* \\
        POLS & -13.25 & -0.04 & -0.95 & 0 & 0 & 0 \\
        PMLS & -13.15 & -0.04 & -0.84 & 0 & 0.07 & 0 \\
    \Xhline{2\arrayrulewidth}
    \end{tabular}
    \caption{Estimates from the additive method, POLS and PMLS. The reported values represent coefficients of each term in the column. All values other than zero are rounded at two decimal points except $xy$ term for the additive method. $xy$ term for the additive method is rounded differently to avoid confusion with the exact zero value.}
    \label{table:lidar_results}
\end{table}

The additive method and PMLS both successfully identified the $y^2$ term, indicating their ability to capture the nonlinear relationship between the variables effectively. However, the same cannot be said for POLS, which failed to identify the $y^2$ term, suggesting redundant intercept terms, one from $\theta_1$ and the other from the mean of $\varsigma$, might have caused some difficulties in estimation procedure. Furthermore, the additive method mis-included $xy$ coefficient in its results, lacking the capability to shrink insignificant variables.
Our proposed method not only accurately estimated regression coefficients, but also outperformed both POLS and the additive method in terms of selection performance. Its ability to select the relevant variables while accurately estimating the parameters offers a better chance for surface estimation in LiDAR-type geodetic applications. 

We notice that the $x$ coefficient survives non-zero in all methods, yet it should have shrunk completed to zero as it was not included in the true surface function \eqref{eq:lidar_curve}. In Figure \ref{fig:experiment}, 
the camera and the surface are designed to face each other, but they could have been misaligned as the experiment was set up manually. We believe this slight misalignment could have led to the inaccurate selection results since the inaccuracy took place over all methods.

\section{Summary and discussion}\label{section:conclusion}

In this paper, we have investigated a modified least squares estimator in nonlinear spatial regression, potentially incorporating a penalty function. We provide the consistency and the asymptotic normality of the proposed estimators. Our simulation study supports that the penalized estimators with LASSO and SCAD penalties work well, even with small sample sizes, and outperform alternative methods. A data example with the LiDAR application also shows that our method demonstrates better performance than existing approaches, making it widely applicable in the field of geodesy. 

We have developed the asymptotic results under increasing domain asymptotic and stationarity. The asymptotic results could be attained under the mixed domain asymptotic with a slightly different convergence order as \cite{lahiri2003central}, a key paper for deriving the asymptotic normality in our study, provides results under the mixed domain asymptotic. Thus, the extension of the asmyptotic results to the mixed domain asymptotic should be straight-forward. While stationarity assumption for the error might be restrictive in real-world applications, we leave exploring results for locally stationary or non-stationary errors in general for future research. Moreover, our study has interesting intersections with the work of \cite{lahiri2006resampling} in that $g(x) \equiv x$ in our results leads to results with $\psi(x)\equiv x$ in \cite{lahiri2006resampling}. Extending our results to encompass M-estimators with nonlinear regression is an intriguing avenue for future investigations.

As in \cite{you2022regularized}, we could take into account a spatial weight matrix in the objective function to improve performance of the estimators. Unfortunately, the lack of necessary knowledge from previous studies did not allow us to obtain the asymptotic properties under the presence of the spatial weight matrices. Nevertheless, we expect estimators with weight matrices to behave at least as well as the proposed estimators without weight matrices as suggested in \cite{you2022regularized}. 

Lastly, we employed the BIC-type criterion to select the tuning parameter in penalty functions. Considering that our method does not involve a likelihood, it would be possible to develop a criterion not depending on the likelihood such as cross validation or generalized cross validation for choosing the tuning parameter and evaluating each model performance.

\section*{Acknowledgements}
Wu was supported by National Science and Technology Council (NSTC) of Taiwan under grant No. 111-2118-M-259-002. The works of Lim, Yoon, and Choi were supported by the National Research Foundation of Korea (NRF) grant funded by the Korea government (MSIT) (No. 2019R1A2C1002213, 2020R1A4A1018207, No.RS-2023-00221762).

\clearpage
\bibliographystyle{chicago}
\bibliography{main}
\clearpage

\newpage
\appendix
\section{Proofs}\label{section:appendix_proof}
We introduce a lemma with reference where the proof of the lemma is provided, so we omit the proof here.
\begin{lemma}[Theorem 3.2 from \cite{lahiri2003central}]\label{lem:clt}
~ \\
A random field $\{\xi(\bm s): \bm s \in \mathcal{R}^d\}$ is stationary with $\E \xi (\bm 0) =0$, $\E |\xi(\bm 0)|^{2+\delta} < \infty$ for some $\delta>0$. In addition, if the conditions below are fulfilled,
\begin{itemize}
    \item [(a)] There exists a function $H_1(\cdot)$ such that for all $\bm h \in \mathcal{R}^d$, as $n \rightarrow \infty$,
    $$\left[\int w_n^2(\eta_n \bm u) \phi(\bm u) d\bm u\right]^{-1}\int w_n(\eta_n \bm u+\bm h)w_n(\eta_n \bm u) \phi^2(\bm u)d\bm u \rightarrow H_1(\bm h)$$
    \item [(b)] $\gamma_{1n}^2 = O(n^a)\ \&\ o\left([\log n]^{-2}\eta_n^{\frac{\tau-d}{4\tau}}\right)$ for some $a \in [0, 1/8)$ and $\tau >d$, where
    $$\gamma_{1n}^2 = \frac{\sup\{|w_n(\bm s)|^2:\bm s\in \eta_n R_0\}}{\E[w_n^2(\eta_n \bm U_1)]}$$
    \item [(c)] $\phi(\bm u)$ is continuous and everywhere positive with support $\bar{R}_0$
    \item [(d)] For $\alpha_1(\cdot)$ from assumption , 
    \begin{itemize}
        \item $\int_1^{\infty} t^{d-1} \alpha_1(t)^{\frac{\delta}{2+\delta}}<\infty$
        \item $\alpha_1(t) \sim t^{-\tau}L(t)$ as $t \rightarrow \infty$ for some slowly varying function $L(\cdot)$
    \end{itemize}
    \item[(e)] For $d\geq 2$ and $\alpha_2(\cdot)$ from assumption , as $t \rightarrow \infty$
    $$\alpha_2(t) = o\left([L(t)^{\frac{\tau+d}{4d\tau}}]^{-1}t^{\frac{\tau-d}{4d}} \right)$$
\end{itemize}
then, with $n/\eta_n^d \rightarrow C_1 \in (0, \infty)$, as $n\rightarrow \infty$
$$\left(n \E[w_n^2(\eta_n \bm U_1)]\right)^{-1/2} \sum_{i=1}^{n} w_n(\bm s_i)\xi(\bm s_i) \overset{d}{\rightarrow} N\left(0, \sigma(\bm 0)+C_1\int \sigma(\bm h) H_1(\bm h) d\bm h\right)$$
\end{lemma}



\newpage

\noindent{\bf Proof of Theorem \ref{theorem:consistency}}

\noindent For a large enough $n$, we have 
\begin{align*}
S_n(\bm \theta_0+a_n\bm v) -S_n(\bm\theta_0) &= a_n\left(\nabla S_n(\bm\theta_0)\right)^{T}\bm{v}
+\frac{1}{2}n^{-1}\bm v^{T}\nabla^2 S_n(\bm{\theta}_0+a_n\bm{v}t) \bm v\\
&:=\mathbb{A}+ \mathbb{B},
\end{align*}
where $t \in (0,1)$. For the term $\mathbb{A}$,
\begin{align*}
\mathbb{A}& = a_n\left(\nabla S_n(\bm\theta_0)\right)^{T}\bm{v}\\
&=-2a_n\bm{v}^{T}\bm{\dot{G}}(\bm\theta_0)^{T}\bm\Sigma_n\bm{\varepsilon},\\
&=-2a_n\bm{v}^{T}\bm{\dot{G}}(\bm\theta_0)^{T}\bm\Sigma_n\bm{\xi},\;\;\mbox{where } \bm{\xi}=\bm{\varepsilon}-\mu\bm{1}
\end{align*}
We follow a similar proof to \cite{you2022regularized} and evaluate $\Var(\mathbb{A})$ since $\mathbb{A}-\E(\mathbb{A}) = O_p\left(\sqrt{\Var(\mathbb{A})}\right)$.
\begin{align*}
    \Var(\mathbb{A}) & = 4 a_n^2 \bm v^T \bm{\dot{G}}(\bm\theta_0)^{T} \bm\Sigma_n \bm \Sigma_\varepsilon \bm\Sigma_n \bm{\dot{G}}(\bm\theta_0) \bm v \\
    & \leq 4a_n^2 \|\bm{\dot{G}}(\bm\theta_0) \bm v\|^2 \lambda_{\max}\left(\bm\Sigma_n \bm\Sigma_\varepsilon \bm\Sigma_n\right) \\
    & = 4a_n^2 \|\bm{\dot{G}}(\bm\theta_0) \bm v\|^2 \lambda_{\max}\left(\bm\Sigma_n \bm\Sigma_\varepsilon\right) \\
    & \leq 4a_n^2 \|\bm{\dot{G}}(\bm\theta_0) \bm v\|^2 \lambda_\varepsilon \\
    & = O\left(na_n^2 \lambda_\varepsilon \right) \|\bm v\|^2,
\end{align*}
where $\lambda_{max}(\bm M)$ refers to the maximum eigenvalue of a matrix $\bm M$. The second equality holds since products of matrices $AB$ and $BA$ share the same non-zero eigenvalues and $\bm\Sigma_n^2 = \bm \Sigma_n$. The second inequality holds because $\bm\Sigma_n$ has eigenvalues of 0 and 1's. The last equality comes from the compactness of the domain $\mathcal{D}\times \Theta$ by Assumption \ref{assumption:general}-(1). Thus, $\Var(\mathbb{A}) = O\left(n a_n^2 \lambda_\varepsilon\right)\|\bm v \|^2$, which leads to
\begin{equation}\label{eq:A}
    \mathbb{A} = O_p\left(n^{1/2} a_n \lambda_\varepsilon^{1/2}\right)\|\bm v \|
\end{equation}
Now, we decompose $\mathbb{B}$ into 4 terms as follows. 
\begin{align*}
    \mathbb{B} & = \frac{1}{2}a_n^2 \bm v^T \nabla^2 \bm{S}_n(\bm \theta_n)\bm v \\
    & = a_n^2 \bm v^T \left(\bm{\dot{G}}(\bm\theta_n)^{T}\bm\Sigma_n \bm{\dot{G}}(\bm\theta_n) - \bm{\dot{G}}(\bm\theta_0)^{T} \bm\Sigma_n \bm{\dot{G}}(\bm\theta_0) \right) \bm v +\bm v ^T \bm{\dot{G}}(\bm\theta_0)^{T} \bm\Sigma_n \bm{\dot{G}}(\bm\theta_0)\bm v \\
    & ~~~ + a_n^2 \bm v^T \bm{\ddot{G}}(\bm\theta_n)^T \left(I\otimes \bm\Sigma_n \bm d(\bm \theta_n, \bm \theta_0) \right) \bm v - a_n^2 \bm v^T \bm{\ddot{G}}(\bm\theta_n)^T \left(I\otimes \bm\Sigma_n \bm\xi  \right)\bm v \\
    & = \mathbb{B}_1+\mathbb{B}_2+\mathbb{B}_3+\mathbb{B}_4
\end{align*}
where $\bm \theta_n = \bm \theta_0 + a_n \bm v t$ and $\bm d(\bm \theta, \bm \theta') = (d_1(\bm\theta, \bm \theta'), \ldots, d_n(\bm\theta, \bm\theta'))^T$ with $d_i(\bm \theta, \bm \theta') = g(\bm x(\bm s_i); \bm \theta)-g(\bm x(\bm s_i); \bm \theta')$. Since $n^{-1}{\bm{\dot{G}}}({\bm\theta})^{T}\bm\Sigma_n {\bm{\dot{G}}}({\bm\theta})$ is a uniformly continuous function of $\bm \theta$ and converges to $\bm \Sigma_g = \bm \Delta_{g^2}-\bm \Delta_g\bm \Delta_g^T$ at $\bm \theta_0$ in probability by Assumptions \ref{assumption:general}-(4) and (5), 
\begin{equation}\label{eq:B1}
    \mathbb{B}_1 = na_n^2\bm v^T\left(n^{-1}\bm{\dot{G}}(\bm\theta_n)^{T}\bm\Sigma_n \bm{\dot{G}}(\bm\theta_n) - n^{-1}\bm{\dot{G}}(\bm\theta_0)^{T} \bm\Sigma_n \bm{\dot{G}}(\bm\theta_0)\right)\bm v = o(na_n^2)\|\bm v\|^2.
\end{equation}
By nature,
$\mathbb{B}_2 = na_n^2 \bm v^T \bm \Sigma_g \bm v (1+o(1))$. Note that $\mathbb{B}_2$ is positive with the order of $na_n^2 = \lambda_\varepsilon$ and $\lambda_\varepsilon$ does not vanish to zero. Next,
\begin{align*}
    \mathbb{B}_3 & = a_n^2 \bm v^T \bm{\ddot{G}}(\bm\theta_n)^T \left(I\otimes \bm\Sigma_n \bm d(\bm \theta_n, \bm \theta_0) \right) \bm v \\
    & = a_n^2 \bm v^T \left[\bm g_{kl}(\bm \theta_n)^T \bm\Sigma_n \bm d(\bm \theta_n, \bm \theta_0) \right]_{k, l = 1, \ldots, p} \bm v,
\end{align*}
where $[a_{kl}]_{k,l=1, \ldots, p}$ is a matrix of which component at $k$-th row and $l$-th column equals to $a_{kl}$. 
\begin{align*}
    \left|\bm g_{kl}(\bm \theta_n)^T \bm\Sigma_n \bm d(\bm \theta_n, \bm \theta_0)\right| & \leq \left|\bm g_{kl}(\bm \theta_n)^T \bm\Sigma_n\bm g_{kl}(\bm \theta_n)\right|^{1/2}\left|\bm d(\bm \theta_n, \bm \theta_0)^T\bm\Sigma_n \bm d(\bm \theta_n, \bm \theta_0) \right|^{1/2} \\
    & \leq \|\bm g_{kl}(\bm \theta_n)\| \|\bm d(\bm \theta_n, \bm \theta_0)\| \\
    & = O(n^{1/2}) O(\|\bm \theta_n - \bm \theta_0\|) \\
    & = O(n^{1/2}a_n)\|\bm v\|
\end{align*}
The first inequality is Cauchy-Schwarz inequality with semi-norm and the second inequality holds since the maximum eigenvalue of $\bm\Sigma_n$ is 1. The compactness of the domain explains the first and second equalities. To sum up, 
\begin{equation}\label{eq:B3}
    \mathbb{B}_3 = O(n^{1/2}a_n^3) \|\bm v\|^3.     
\end{equation}
Similarly, to deal with $\mathbb{B}_4$ we evaluate $\left[\bm g_{kl}(\bm \theta_n)^T \bm\Sigma_n \bm \xi \right]$ for all $k,l =1,\ldots, p$.
\begin{align*}
    \Var\left(\bm g_{kl}(\bm \theta_n)^T \bm\Sigma_n \bm \xi\right) & = \bm g_{kl}(\bm \theta_n)^T \bm\Sigma_n \bm \Sigma_\varepsilon \bm\Sigma_n \bm g_{kl}(\bm \theta_n) \\
    & \leq \left\| \bm g_{kl}(\bm \theta_n)\right\|^2 \lambda_\varepsilon \\
    & = O(n\lambda_\varepsilon).
\end{align*}
Recall that the maximum eigenvalue of $\bm\Sigma_n \bm\Sigma_\varepsilon\bm\Sigma_n$ is not greater than $\lambda_\varepsilon$. Since $\E(\bm g_{kl}(\bm \theta_n)^T \bm\Sigma_n \bm \xi)=0$, $\bm g_{kl}(\bm \theta_n)^T \bm\Sigma_n \bm \xi=O_p(n^{1/2}\lambda_\varepsilon^{1/2})$ and
\begin{equation}\label{eq:B4}
    \mathbb{B}_4 = O_p(n^{1/2}a_n^2\lambda_\varepsilon^{1/2})\|\bm v\|^2.
\end{equation}
Therefore, conditional on $\bm U$,
\begin{align}
    \mathbb{B} & = \mathbb{B}_1 +\mathbb{B}_2+\mathbb{B}_3+\mathbb{B}_4 \n \\
    & = o(na_n^2)\|\bm v\|^2 +na_n^2 \bm v^T \bm \Sigma_g \bm v (1+o(1))+O(n^{1/2}a_n^3)\|\bm v\|^3 +O_p(n^{1/2}a_n^2 \lambda_\varepsilon^{1/2})\|\bm v\|^2 \n\\
    & = na_n^2 \bm v^T\bm\Sigma_g \bm v (1+o(1)) + O_p(n^{1/2}a_n^2 \lambda_\varepsilon^{1/2})\|\bm v\|^2 \n \\
    & = na_n^2\bm v^T\bm\Sigma_g \bm v (1+o_p(1))  \label{eq:B}
\end{align}
The last equality holds for $\lambda_\varepsilon =o(n)$ in Assumption \ref{assumption:stationary}-(1). Finally, by \eqref{eq:A} and \eqref{eq:B}, we attain
\begin{align}
    \mathbb{A+B} & = O_p\left(n^{1/2}a_n \lambda_\varepsilon^{1/2} \right)\|\bm v\| + na_n^2\bm v^T\bm\Sigma_g \bm v (1+o_p(1)) \n \\
    & = O_p\left(\lambda_\varepsilon \right)\|\bm v\| + \lambda_\varepsilon\bm v^T\bm\Sigma_g \bm v (1+o_p(1)) \label{eq:AB}
\end{align}
With large enough $\|\bm v\|$, \eqref{eq:AB} stays positive with probability tending to 1, which completes the proof.

\clearpage

\noindent{{\bf Proof of Theorem ~\ref{theorem:consistency-penalty}}}\\

Let $b_n = a_n+c_n$, 
\begin{eqnarray}\label{eq:thm2}
	&& Q_n(\bm\theta_0+b_n \bm v)-Q_n(\bm\theta_0) \n \\
	& & ~~~ = \mathbb{A'+B'}+\displaystyle n \left(\sum_{i=1}^{p} p_{\lambda_{n}}(|\theta_{0i}+b_n v_i|)-p_{\lambda_{n}} (|\theta_{0i}|) \right) \n \\
	& & ~~~  =\mathbb{A'+B'}+\displaystyle n \left(\sum_{i=1}^{s} p_{\lambda_{n}}(|\theta_{0i}+b_n v_i|)-p_{\lambda_{n}} (|\theta_{0i}|) \right)+n\sum_{i=s+1}^{p} p_{\lambda_{n}}(|\theta_{0i}+b_n v_i|) \n \\
	& & ~~~ \geq \mathbb{A'+B'} +n\left(\sum_{i=1}^{s} q_{\lambda_{n}}(|\theta_{0i}|)sgn(\theta_{0i})b_n v_i +\frac{1}{2}q_{\lambda_n}'(|\theta_{0i}|)b_n^2 v_i^2(1+o(1))\right) \n \\
	& & ~~~ \geq \mathbb{A'+B'+C+D},
\end{eqnarray}
where $\mathbb{A}'$ and $\mathbb{B}'$ are defined similarly to $\mathbb{A}$ and $\mathbb{B}$ in the proof of Theorem \ref{theorem:consistency} with replacement of $a_n$ to $b_n$. $\mathbb{C}=n\sum_{i=1}^{s} q_{\lambda_{n}}(|\theta_{0i}|)sgn(\theta_{0i})b_n v_i$ and $\mathbb{D}=(n/2)\sum_{i=1}^{s}q_{\lambda_n}'(|\theta_{0i}|)b_n^2 v_i^2(1+o(1))$. The expression of $\mathbb{D}$ comes from Assumption \ref{assumption:penalty}-(2). Then, by \eqref{eq:A} and \eqref{eq:B}, $$\mathbb{A}'+\mathbb{B}' = O_p\left(n^{1/2}b_n\lambda_\varepsilon^{1/2}\right)\|\bm v\|+nb_n^2\bm v^T \bm \Sigma_g \bm v (1+o_p(1))$$
For the desired result, it is enough to show that $\mathbb{C}=O(nb_n^2)\|\bm v\|$ and $\mathbb{D}=o(nb_n^2)\|\bm v\|^2$. By the definition of $b_n$ and Assumption \ref{assumption:penalty}-(1), 
\begin{equation}\label{eq:C}
    |\mathbb{C}| \leq n\sum_{i=1}^{s}|q_{\lambda_{n}}(|\theta_{0i}|)b_n v_i| \leq n b_nc_n \sum_{i=1}^{s} |v_i| \leq  nb_n^2\sqrt{s}\|\bm v\| \n
\end{equation}
\begin{equation}\label{eq:D}
    |\mathbb{D}| \leq \displaystyle \max_{1\leq i\leq s}|q_{\lambda_n}'(|\theta_{0i}|)| n b_n^2 \| \bm v\|^2 = o(nb_n^2) \|\bm v\|^2  \n
\end{equation}
Putting the results above all together, we obtain $\mathbb{B}'$ dominates $\mathbb{C}$ and $\mathbb{D}$ with large enough $\|\bm v\|$. By Assumption \ref{assumption:penalty}-(1), $b_n=O(n^{-1/2}\lambda_\varepsilon^{1/2})$, so with the probability tending to 1, \eqref{eq:thm2} is positive.

\clearpage

\noindent{\bf{Proof of Theorem ~\ref{theorem:normality}}} \\

\noindent By the mean-value theorem of a vector-valued function \citep{feng2013mvt},
\begin{align*}
	 & n^{-1/2}\nabla S_n(\bm\theta_0) \\
	 & \hspace{.5cm} = n^{-1/2}\left(\nabla S_n\left(\hat{\bm\theta}^{(s)}\right)+\left(\int^{1}_{0}\nabla^2 S_n\left(\bm{\hat\theta}^{(s)}+\left(\bm\theta_0-\hat{\bm\theta}^{(s)}\right)t\right)dt \right)^{T}\left(\bm\theta_0-\hat{\bm\theta}^{(s)}\right)\right)\\
	 & \hspace{.5cm} = n^{-1/2}\left(\int^{1}_{0}\nabla^2 S_n\left(\bm{\hat\theta}^{(s)}+\left(\bm\theta_0-\hat{\bm\theta}^{(s)}\right)t\right)dt \right)^{T}\left(\bm\theta_0-\hat{\bm\theta}^{(s)}\right),
\end{align*}
For a arbitrary vector $\bm v \in \mathcal{R}^d$, we consider $n^{-1/2} \bm v^T \nabla S_n(\bm\theta_0)$ as follows.
$$n^{-1/2} \bm v^T \nabla S_n(\bm\theta_0)=-2n^{-1/2}\bm{v}^T\bm{\dot{G}}(\bm\theta_0)^T \bm \xi +2n^{-1/2}\bm v^T \bm{\dot{G}}(\bm\theta_0)^{T}\left(\frac{1}{n}\bm 1 \bm 1^T\right)\bm \xi.$$
Since $\bm x(\bm s_i) = \bm x(\eta_n\bm u_i)$ is independent, by WLLN $n^{-1}\bm{v}^{T}\bm{\dot{G}}(\bm\theta_0)^{T}\bm{1}=\bm v^T \bm \Delta_g (1+o_p(1))$. Thus, conditional on $\bm U$,
\begin{align*}
    n^{-1/2} \bm v^T \nabla S_n(\bm\theta_0) & = -2n^{-1/2}\bm{v}^{T}\bm{\dot{G}}(\bm\theta_0)^{T} \bm \xi + 2n^{-1/2}\bm v^T \bm \Delta_g  \bm{1}^T\bm \xi+o_p(1)\\
    & = -2n^{-1/2}\sum_{i=1}^{n} \bm{v}^{T}\left(\frac{\partial g(\bm x(\bm s_i);\bm\theta_0)}{\partial \bm\theta}-\bm \Delta_g \right)\xi(\bm s_i) +o_p(1)
\end{align*}
By Assumptions \ref{assumption:general}, \ref{assumption:stationary} and the lemma \ref{lem:clt} with $w_n(\bm s_i) = \bm{v}^{T}\left(\frac{\partial g(\bm x(\bm s_i);\bm\theta_0)}{\partial \bm\theta}-\bm \Delta_g \right)$, 
\begin{equation}
    -2n^{-1/2} \sum_{i=1}^{n} w_n(\bm s_i) \xi(\bm s_i) \overset{d}{\rightarrow} N\left(0, 4\bm v^{T}\left(\sigma(\bm 0) \bm \Sigma_g + \int \sigma(\bm h) \bm r(\bm h) d\bm h \right)\bm v \right)
\end{equation}
where $\bm \Sigma_g = \bm \Delta_{g^2}-\bm \Delta_g\bm \Delta_g^T,\ \bm r(\bm h) = \bm r_2(\bm h)-\bm \Delta_g(\bm r_1(\bm 0)+\bm r_1(\bm h))^T+\E[\phi(\bm U_1)]\bm \Delta_g\bm \Delta_g^T$, and $\sigma(\bm h) = \E\left[\xi(\bm 0)\xi(\bm h)\right]$. The variance expression follows from the lemma 1 since 
\begin{align*}
    \int w_n^2(\eta_n \bm u) \phi(\bm u)d\bm u & \overset{p}{\rightarrow} \bm v^{T}\bm \Sigma_g\bm v    \\
    \int w_n(\eta_n \bm u+\bm h)w_n(\eta_n \bm u) \phi^2(\bm u)d \bm u & \overset{p}{\rightarrow} \bm v^{T}\left(\bm r_2(\bm h)-\bm \Delta_g(\bm r_1(\bm 0)+\bm r_1(\bm h))^T+\E[\phi(\bm U_1)]\bm \Delta_g\bm \Delta_g^T\right)\bm v
\end{align*}
Therefore, by the conditional version of Cramer-Wold device,
\begin{equation}\label{eq:normality}
n^{-1/2} \nabla S_n(\bm \theta_0) \overset{d}{\longrightarrow} N\left(0, 4
\bm \Sigma_\theta\right)
\end{equation}
where $\bm \Sigma_\theta=\sigma(0)\bm \Sigma_g  + \int \sigma(\bm h) \bm r(\bm h) d\bm h$. The proof is completed by Slutsky's theorem if we attain 
\begin{equation}\label{eq:remainder}
    n^{-1}\left(\int^{1}_{0}\nabla^2 S_n\left(\bm{\hat\theta}^{(s)}+(\bm\theta_0-\hat{\bm\theta}^{(s)})t\right)dt \right) \overset{p}{\longrightarrow} 2\bm \Sigma_g
\end{equation}
where $\bm\Sigma_g = \bm \Delta_{g^2}-\bm \Delta_{g}\bm\Delta_{g}^T$. For \eqref{eq:remainder}, it is enough to show that
\begin{align}
    & n^{-1}\nabla^2 S_n(\bm \theta_0) \overset{p}{\rightarrow} 2\bm\Sigma_g, \label{eq:remainder_1}\\
    & n^{-1/2} \left(\int^{1}_{0}\nabla^2 S_n\left(\bm{\hat\theta}^{(s)}+(\bm\theta_0-\hat{\bm\theta}^{(s)})t\right)dt -\nabla^2 S_n(\bm \theta_0) \right) =o_p(1). \label{eq:remainder_2}
\end{align}
For \eqref{eq:remainder_1},
$$n^{-1} \nabla^2 S_n(\bm \theta_0) = 2n^{-1}\dot{\bm G}(\bm\theta_0)^T\bm\Sigma_n \dot{\bm G}(\bm\theta_0)-2n^{-1}\ddot{\bm G}(\bm\theta_0)^T(I \otimes \bm \Sigma_n \bm \xi).$$
The first term converges to $2\bm\Sigma_g$ by Assumptions \ref{assumption:general}-(4) and (5) and the second term is $O_p(n^{-1/2}\lambda_\varepsilon^{1/2})$ by similar procedure to \eqref{eq:B4}. Thus, by Assumption \ref{assumption:stationary}-(1), we achieve \eqref{eq:remainder_1}. Next, since $\hat{\bm\theta}^{(s)}$ is root-n-consistent under Assumption \ref{assumption:stationary}-(2),
\begin{align*}
    & n^{-1}\left\|\int^{1}_{0}\nabla^2 S_n\left(\bm{\hat\theta}^{(s)} +  (\bm\theta_0-\hat{\bm\theta}^{(s)})t\right)dt-\nabla^2 S_n(\bm{\theta_0})\right\| \\
     & ~~~~~~~~~~~~~~~~~~~~~~~~~~~ \leq n^{-1}\max_{\tilde{\bm\theta}\in B(\bm\theta_0,n^{-1/2} C)}\left\| \nabla^2 S_n(\tilde{\bm\theta})- \nabla^2 S_n(\bm\theta_0)\right\|.
\end{align*}
The above term can be decomposed into 3 terms as follows.
\begin{align*}
    & n^{-1}\max_{\tilde{\bm\theta}\in B(\bm\theta_0,n^{-1/2} C)}\left\| \nabla^2 S_n(\tilde{\bm\theta})- \nabla^2 S_n(\bm\theta_0)\right\| \\
    & \hspace{2cm} \leq 2n^{-1}\max_{\tilde{\bm\theta}\in B(\bm\theta_0,n^{-1/2} C)}\left\|\dot{\bm G}(\tilde{\bm\theta})^T\bm\Sigma_n \dot{\bm G}(\tilde{\bm\theta})-\dot{\bm G}(\bm\theta_0)^T\bm\Sigma_n \dot{\bm G}(\bm\theta_0) \right\| \\
    & \hspace{2cm} + 2n^{-1}\max_{\tilde{\bm\theta}\in B(\bm\theta_0,n^{-1/2} C)}\left\| \ddot{\bm G}(\tilde{\bm\theta})^T(I\otimes \bm\Sigma_n \bm d(\tilde{\bm\theta},\bm\theta_0)) \right\| \\
    & \hspace{2cm} + 2n^{-1}\max_{\tilde{\bm\theta}\in B(\bm\theta_0,n^{-1/2} C)}\left\| \left(\ddot{\bm G} (\tilde{\bm\theta})-\ddot{\bm G}(\bm\theta_0)\right)^T (I \otimes \bm\Sigma_n \bm \xi) \right\|.
\end{align*}
The first term is $o(1)$ conditional on $\bm U$ by similar procedure to \eqref{eq:B1}. The second part converges to 0 by
\begin{align*}
    2n^{-1}\left\| \ddot{\bm G}(\tilde{\bm\theta})^T\left(I \otimes\bm\Sigma_n\bm d(\tilde{\bm\theta}, \bm \theta_0)\right)\right\|&= 2n^{-1}\left\|\left[\bm g_{kl}(\tilde{\bm\theta})^T\bm\Sigma_n \bm d(\tilde{\bm\theta}, \bm\theta_0)\right]_{k,l=\{1,\ldots,p\}}\right\|  \\
    & \leq 2n^{-1}O\left(\max_{k,l}\left|\bm g_{kl}(\tilde{\bm\theta})^T\bm\Sigma_n \bm d(\tilde{\bm\theta}, \bm\theta_0)\right|\right) \\
    & \leq 2n^{-1}O\left(\max_{k,l}\left\|\bm g_{kl}(\tilde{\bm\theta}) \right\|\left\| \bm d(\tilde{\bm\theta}, \bm\theta_0)\right\|\right)\\
    & \leq 2n^{-1} O\left(n^{1/2}\|\tilde{\bm\theta}-\bm\theta_0\| \right) \\
    & = O\left(n^{-1}\right).
\end{align*}
The first and third inequalities hold since $p$ is finite and $g$ is bounded, respectively. For the last term, similarly to \eqref{eq:B4} we evaluate the variance of $(k,l)$-th component.
\begin{align*}
    \Var\left(\left(\bm g_{kl}(\tilde{\bm\theta}) -\bm g_{kl}(\bm\theta_0)\right)^T\bm\Sigma_n \bm \xi\right) & = \left(\bm g_{kl}(\tilde{\bm\theta}) -\bm g_{kl}(\bm\theta_0)\right)^T\bm\Sigma_n \bm\Sigma_\varepsilon\bm\Sigma_n \left(\bm g_{kl}(\tilde{\bm\theta}) -\bm g_{kl}(\bm\theta_0)\right) \\ 
    & \leq \lambda_\varepsilon\left\|\bm g_{kl}(\tilde{\bm\theta}) -\bm g_{kl}(\bm\theta_0)\right\|^2 \\
    & \leq \lambda_\varepsilon \|\tilde{\bm\theta}-\bm\theta_0\|^2 \\
    & = O\left(n^{-1}\lambda_\varepsilon\right) = O\left(n^{-1}\right)
\end{align*}
This indicates that the last term is $O_p(n^{-1/2})$. To sum up, 
$$n^{-1}\max_{\tilde{\bm\theta}\in B(\bm\theta_0,n^{-1/2} C)}\left\| \nabla^2 S_n(\tilde{\bm\theta})- \nabla^2 S_n(\bm\theta_0)\right\| \leq o(1)+O\left(n^{-1}\right) + O_p\left(n^{-1/2}\right)=o_p(1).$$
At last, we attain \eqref{eq:remainder_2} and, therefore, \eqref{eq:remainder}. Bringing \eqref{eq:normality} and \eqref{eq:remainder} together with Slutsky's theorem, we obtain
$$n^{1/2}\left(\hat{\bm\theta}^{(s)}-\bm\theta_0\right) \overset{d}{\longrightarrow} N\left(0, \bm\Sigma_g^{-1}\bm\Sigma_\theta\bm\Sigma_g^{-1}  \right).$$

\clearpage
\noindent {\bf{Proof of Theorem ~\ref{theorem:sparsity}}}\\


\noindent Proof of (i)\\
    We show for $\epsilon>0\ $\&$\ i \in \{s+1, \ldots, p\},$ $P(\hat{\theta}_i \neq 0) <\epsilon$.
	First, we break $(\hat{\theta}_i\neq 0)$ into two sets: 	
	\begin{align*}
	\{\hat{\theta}_i\neq 0\} & = \left\{\hat{\theta}_i\neq 0, |\hat{\theta}_i|\geq C n^{-1/2}\right\}+\left\{\hat{\theta}_i\neq 0, |\hat{\theta}_i|< C n^{-1/2}\right\}\\
	& := E_n+F_n.
	\end{align*}
	Then, it is enough to show for any $\epsilon > 0$, $P(E_n) <\epsilon/2 $ and $P(F_n) < \epsilon/2$.
	For any $\epsilon >0$, we can show $P(E_n)<\epsilon/2$ for large enough $n$ because  $\|\bm{\hat\theta}-\bm\theta_0\|=O_p(n^{-1/2})$ under Assumption \ref{assumption:stationary}-(2).	
	To verify $P(F_n)<\epsilon/2$ for large enough $n$, we first show $n^{1/2} q_{\lambda_{n}}(|\hat\theta_i|)=O_p(1)$ on the set $F_n$. By the mean-value theorem again,
	$$n^{-1/2}\nabla S_n(\bm\theta_0)
	= n^{-1/2}\left(\nabla S_n({\bm\theta})+\left(\int^{1}_{0}\nabla^2 S_n\left(\bm\theta+(\bm\theta_0-{\bm\theta})t\right)dt \right)^{T}(\bm\theta_0-{\bm\theta})\right).$$
	Recall that $n^{-1/2}\nabla S_n(\bm\theta_0) = O_p(1)$ from \eqref{eq:normality}. In addition, with $\|\bm\theta_0-\bm\theta\|=O(n^{-1/2})$, the similar results of \eqref{eq:remainder} inform that
	\begin{equation}\label{eq:secondderivative}
	\left\|n^{-1/2}\left(\int^{1}_{0}\nabla^2 S_n\left(\bm\theta+(\bm\theta_0-{\bm\theta})t\right)dt \right)^{T}(\bm\theta_0-{\bm\theta})\right\| = O_p(1).
	\end{equation}
	Then, we have
$$\left\|n^{-1/2}\nabla S_n(\bm \theta)\right\|=O_p(1)\;\hbox{for $\bm\theta$ satisfying $\|\bm\theta-\bm\theta_0\|\leq Cn^{-1/2}$.}$$
Since $\hat{\bm\theta}$ is the local minimizer of $Q_n(\bm\theta)$ with $\|\hat{\bm\theta}-\bm\theta_0\|=O_p(n^{-1/2})$, we attain 
$$n^{1/2} q_{\lambda_{n}}(|\hat{\theta}_i|)=O_p(1)$$
from $$n^{-1/2}\left.\frac{\partial Q_n(\bm \theta)}{\partial\theta_i}\right|_{\bm\theta=\hat{\bm\theta}}=n^{-1/2}\left.\frac{\partial S_n(\bm \theta)}{\partial\theta_i}\right|_{\bm\theta=\hat{\bm\theta}}+n^{1/2} q_{\lambda_{n}}(|\hat{\theta_i}|)sgn(\hat\theta_i).$$ Therefore, there exists $M^\prime$ such that $P\{n^{1/2} q_{\lambda_n}(|\hat{\theta}_i|)>M^{\prime}\}<\epsilon/2$ for large enough $n$, which implies
$P\{\hat{\theta}_i\neq 0, |\hat{\theta}_i|<Cn^{-1/2},  n^{1/2} q_{\lambda_n}(|\hat{\theta}_i|)>M^{\prime}\}<\epsilon/2$. At last, by Assumptions \ref{assumption:penalty}-(3) and (4), for large enough $n$,
$$\{\hat{\theta}_i\neq 0, |\hat{\theta}_i|<C n^{-1/2},  n^{1/2} q_{\lambda_n}(|\hat{\theta}_i|)>M^{\prime}\}=\{\hat{\theta}_i\neq 0, |\hat{\theta}_i|<C n^{-1/2}\},$$ 
which leads to $P(F_n)<\epsilon/2$. \\

\noindent Proof of (ii)\\
By Taylor expansion,
	\begin{align*}	
		n^{-1/2}\nabla Q_n(\hat{\bm\theta}) & = n^{-1/2}\nabla S_n(\hat{\bm\theta})+n^{1/2}{\bm{q}}_{\lambda_n}(\hat{\bm\theta})\cdot sgn(\hat{\bm\theta})\\
		&=n^{-1/2}\left(\nabla S_n(\bm\theta_0)+\left(\int^{1}_{0}\nabla^2 S_n\left(\bm{\theta}_0+(\bm{\hat\theta}-\bm\theta_0)t\right) \right)^{T}(\bm{\hat\theta}-\bm\theta_0)\right)\\
		&+n^{1/2}\bm{q}_{\lambda_n}(\hat{\bm\theta})\cdot sgn(\hat{\bm\theta})
	\end{align*} 
	where $\bm{q}_{\lambda_n}(\hat{\bm\theta}) \cdot sgn(\hat{\bm\theta})=\left(q_{\lambda_n}(|\hat{\theta}_1|)sgn(\hat\theta_1),..,q_{\lambda_n}(|\hat{\theta}_p|)sgn(\hat\theta_p)\right)^{T}.$ Since $\hat{\bm\theta}$ is the local minimizer of $Q_n({\bm\theta})$,
	$\nabla Q_n(\hat{\bm\theta})=0$, which implies
	$$-n^{-1/2}\nabla S_n({\bm\theta_0})=n^{-1}\left(\int^{1}_{0}\nabla^2 S_n\left(\bm{\theta}_0+(\bm{\hat\theta}-\bm\theta_0)t\right)dt \right)^{T}\left(n^{1/2}(\hat{\bm\theta}-\bm\theta_0)\right)+n^{1/2}\bm{q}_{\lambda_n}(\hat{\bm\theta})\cdot sgn(\hat{\bm\theta}).$$
	Since $n^{-1/2}\nabla S_n(\bm\theta_0) \overset{d}{\rightarrow}N(4\bm\Sigma_\theta)$ by \eqref{eq:normality} and $n^{-1}\left(\int^{1}_{0}\nabla^2 S_n(\bm\theta_0+(\bm{\hat\theta}-\bm\theta_0)t)dt \right)\overset{p}{\rightarrow} 2\bm\Sigma_g$ by the similar results of \eqref{eq:remainder},
	$$n^{1/2}\left(\hat{\bf{\bm\theta}}-\bm\theta_{0}+(2\bm\Sigma_g)^{-1}\bm q_{\lambda_n}(\hat{\bm\theta})sgn(\hat{\bm\theta})\right)~ \stackrel{d}{\rightarrow}N\left(0,\bm \Sigma_g^{-1}\bm \Sigma_\theta\bm \Sigma_g^{-1}\right).$$
	Finally,
	$$n^{1/2}\left({\hat{\bf{\bm\theta}}}_{11}-\bm\theta_{01}+(2\bm \Sigma_g)_{11}^{-1}\bm \beta_{n, s}\right)~ \stackrel{d}{\rightarrow}N\left(0,\left(\bm \Sigma_g^{-1}\bm \Sigma_{\theta}\bm \Sigma_g^{-1}\right)_{11}\right),$$
	where $(\bm \Sigma_g)_{11}^{-1}$ is the $s\times s$ upper-left matrix of $\bm\Sigma_g^{-1}$.

\end{document}